\documentclass[pra, amsmath, amssymb, superscriptaddress,reprint]{revtex4-1}
\usepackage{pstricks}
\usepackage{color}
\usepackage{amsmath,amssymb}
\usepackage{epsfig}
\usepackage{graphicx}
\usepackage{dcolumn}
\usepackage{bbold}
\usepackage{natbib}
\usepackage{datetime}
\usepackage{textcomp}
\usepackage{bm}
\usepackage{color}
\usepackage{amsfonts}
\usepackage{lineno}
\usepackage{blindtext}
\usepackage[utf8]{inputenc}
\usepackage{float}
\usepackage{tabularx}
\usepackage{physics}
\usepackage{braket}
\usepackage{docmute}
\usepackage[colorlinks=true,citecolor={red},urlcolor={black},linkcolor={red}]{hyperref}

\usepackage{upgreek}
\begin{document}

\title{Implementation of a time-dependent multiconfiguration
self-consistent-field method for coupled
electron-nuclear dynamics in diatomic molecules driven by intense laser pulses} 
\author{Yang Li}
\email{yangli@atto.t.u-tokyo.ac.jp}
\affiliation{Graduate School of Engineering, The University of Tokyo, 7-3-1 Hongo, Bunkyo-ku, Tokyo 113-8656, Japan}
%%%%%%%%%%%%%%%%%%%%%%%%%
\author{Takeshi Sato}
\email{sato@atto.t.u-tokyo.ac.jp}
\affiliation{Graduate School of Engineering, The University of Tokyo, 7-3-1 Hongo, Bunkyo-ku, Tokyo 113-8656, Japan}
\affiliation{Photon Science Center, School of Engineering, University of Tokyo, 7-3-1 Hongo, Bunkyo-ku, Tokyo 113-8656, Japan}
\affiliation{Research Institute for Photon Science and Laser Technology, University of Tokyo, 7-3-1 Hongo, Bunkyo-ku, Tokyo 113-0033, Japan}
%%%%%%%%%%%%%%%%%%%%%%%%%
\author{Kenichi L. Ishikawa}
\email{ishiken@n.t.u-tokyo.ac.jp}
\affiliation{Graduate School of Engineering, The University of Tokyo, 7-3-1 Hongo, Bunkyo-ku, Tokyo 113-8656, Japan}
\affiliation{Photon Science Center, School of Engineering, University of Tokyo, 7-3-1 Hongo, Bunkyo-ku, Tokyo 113-8656, Japan}
\affiliation{Research Institute for Photon Science and Laser Technology, University of Tokyo, 7-3-1 Hongo, Bunkyo-ku, Tokyo 113-0033, Japan}
%%%%%%%%%%%%%%%%%%%%%%%%%%%%%%%%%%%%%%%
\begin{abstract}
We present an implementation of a time-dependent multiconfiguration self-consistent-field method [R. Anzaki, T. Sato and K. L. Ishikawa, \href{https://doi.org/10.1039/C7CP02086D}{Phys. Chem. Chem.
Phys. \textbf{19}, 22008 (2017)}] with the full configuration-interaction expansion for coupled electron-nuclear dynamics in diatomic molecules subject to a strong laser field. 
In this method, the total wave function is expressed as a superposition of different configurations constructed from time-dependent electronic Slater determinants and time-dependent orthonormal nuclear basis functions. 
The primitive basis functions of nuclei and electrons are strictly independent of each other without invoking the Born-Oppenheimer approximation. 
Our implementation treats the electronic motion in its full dimensionality on curvilinear coordinates, while the nuclear wave function is propagated on a one-dimensional stretching coordinate with rotational nuclear motion neglected. 
We apply the present implementation to high-harmonic generation and dissociative ionization of a hydrogen molecule and discuss the role of electron-nuclear correlation.
\end{abstract}
%%%%%%%%%%%%%%%%%%%%%%%%%%%%%%%%%%%%%%%
%\pacs{31.15.X, 32.80.Rm, 33.80.Rv, 42.65.Ky}
\maketitle
%%%%%%%%%%%%%%%%%%%%%%%%%%%%%%%%%%%%%%%%%%%%%%%%%%%%%%
\section{Introduction}\label{sec1}
The rapid development of laser technology  enables the generation of coherent light sources that cover a broad
frequency range, from the mid-infrared, to the x-ray regime \cite{hentschel2001attosecond,paul2001observation,chini2014generation,pupeza2015high,popmintchev2012bright}. The emergence of these new light sources has opened up new opportunities for many scientific fields, such as ultrafast atomic and molecular physics \cite{Agostini2004,Krausz2009,corkum2007attosecond,Gallmann2012}, attosecond chemistry \cite{calegari2014ultrafast,nisoli2017attosecond}, and material science \cite{luu2015extreme,hohenleutner2015real,ghimire2011observation,uzan2020attosecond}, with a goal to visualize, understand and ultimately control electron and nuclear dynamics in atoms, molecules, and solids. 
In particular, when molecules are subject to ultrashort intense laser pulses, the induced molecular dynamics is the simultaneous motion of atomic nuclei and electrons forming a molecular entity \cite{lepine2014attosecond,ranitovic2014attosecond}. An accurate description of such processes requires treating both electronic and nuclear degrees of freedom on an equal footing. At present, \textit{ab initio} theoretical description of coupled electron-nuclear dynamics in molecules remains a challenge.

For the investigation of multielectron dynamics in intense laser fields, the multiconfiguration time-dependent Hartree-Fock (MCTDHF) method has been developed \cite{zanghellini2003mctdhf,kato2004time,caillat2005correlated,nest2005multiconfiguration}. In this approach, the time-dependent total electronic wave function $\Psi(t)$ is given in the configuration-interaction (CI) expansion,
\begin{equation}\label{eq:total_ewfun}
    \Psi(\mathbf{x}_1,\cdots,\mathbf{x}_n,t) = \sum_I C_I(t)\Phi_I(\mathbf{x}_1,\cdots,\mathbf{x}_n,t),
\end{equation}
with $\mathbf{x}=\{\mathbf{r},\sigma\}$ being the spatial-spin variable of the electron. The electronic configuration $\Phi_I(\mathbf{x}_1,\cdots,\mathbf{x}_n,t)$ is a Slater determinant constructed from spin orbital functions $\{\psi(\mathbf{r},t)\times s(\sigma)\}$, where $\{\psi(\mathbf{r},t)\}$ and $\{ s(\sigma)\}$ are electronic spatial orbitals and spin functions, respectively. Both CI coefficients $\{C_I\}$ and orbitals are simultaneously propagated in time, which provides an adequate representation of the temporal change of the wave function with a reduced number of determinants compared to fixed orbital treatment. In the MCTDHF method, the summation in Eq.~(\ref{eq:total_ewfun}) runs over all possible configurations for a given set of spin orbitals, which are conventionally referred to as full-CI expansion. In the full-CI case, the computational cost due to CI coefficients is proportional to the number of electron distributions among all spin orbitals, which increases factorially with the number of electrons, thus hindering
its application beyond small molecular systems. To subjugate this difficulty, variants that go beyond the restriction to the full-CI expansion are now under active development, which we refer to as time-dependent multiconfiguration self-consistent-field (TD-MCSCF) methods \cite{Ishikawa2015,lode2020colloquium} in the general case. Representative examples include the time-dependent complete-active-space self-consistent-field (TD-CASSCF) method \cite{Sato2013,Sato2016,Orimo2018}, the time-dependent occupation-restricted multiple active-space (TD-ORMAS) method \cite{Sato2015,orimo2019application}, and the time-dependent restricted-active-space self-consistent-field (TD-RASSCF) method \cite{Haxton2015,miyagi2013time,Miyagi2014}. These methods are based on the truncated CI expansion within the chosen active orbital space, which are computationally more efficient and at the same time provide compact representation of multielectron dynamics in a given physical condition without sacrificing accuracy.

Efforts have been made to describe the coupled electronic and nuclear dynamics of molecules within the framework of the TD-MCSCF method by multiple groups and from diverse perspectives. Nest developed
the multiconfiguration electron-nuclear dynamics (MCEND) method \cite{nest2009multi,ulusoy2012multi}, where the molecular wave function is represented as a sum over products of determinants for the electrons and Hartree products for the nuclei. By introducing a set of atomic orbitals on "ghost" centers along the internuclear axis of diatomic systems, this method is computationally effective to investigate both electronic and vibrational photo-excitations. However, the MCEND method fails to describe photoionization and dissociation processes due to the use of Gaussian bases functions as electronic basis and limited grid range of nuclear bases \cite{aebersold2019coupled}. These two processes are keys in understanding the strong coupling of electronic and nuclear dynamics to strong laser fields. Kato and Yamanouchi developed an extended MCTDHF method \cite{kato2009time} by using a basis of Slater determinants with time-dependent nuclear orbitals for the fermionic protons while keeping heavy nuclei fixed. This method has been applied to calculate the ground-state properties of a model one-dimensional hydrogen molecule \cite{ide2014non}, which shows good agreement with the direct solution of the time-dependent Schrödinger equation (TDSE). Very recently, this method has also been adopted for real-time simulation of hydrogen molecular ions (with one electron and one nuclear degree of freedom) irradiated by an intense laser pulse \cite{lotstedt2019time}. In another work, Haxton and McCurdy developed a promising way to include quantum treatment of nuclei into MCTDHF for the special case of diatomic molecules. By introducing prolate spheroidal coordinates for the electronic wave function \cite{haxton2011multiconfiguration}, the total wave function is expanded in terms of configurations of orbitals which depend on the internuclear distance. Thus, the trivial but strong correlation between the positions of the nuclei and electrons due to their Coulomb attraction is naturally included and an efficient procedure for evaluating overlap and Hamiltonian matrix elements is formulated. On one hand, geometry-dependent electron orbitals satisfy the key physical condition for the inclusion of strong electron-nuclear coupling. It is demonstrated that such treatment provides a rapidly convergent representation of the vibrational states of diatomics such as $\mathrm{HD^+}$, $\mathrm{H_2}$, $\mathrm{HD}$, and $\mathrm{LiH}$. On the other hand, the parametric dependence of electron orbitals on the nuclear geometry causes qualitative failure in calculating the dissociative photoionization cross section of $\mathrm{H_2^+}$ \cite{haxton2015qualitative}. Some alternatives of the prolate spheroidal coordinates have been suggested to resolve this issue.
In a series of papers \cite{alon2007multiconfigurational,alon2007unified,alon2009many,alon2012recursive}, Alon and Cederbaum have formulated a multiconfiguration method to describe systems consisting of different types of identical particles with particle conversion taken into account. Stepping forward in this direction, a fully general TD-MCSCF method \cite{anzaki2017fully} has been developed in our group, which can describe the dynamics of a many-body system comprising any arbitrary kinds and numbers of fermions and bosons subject to an external field. The equations of motion (EOMs) of CI-coefficients and spin orbitals are derived for general configuration spaces, based on the time-dependent variational principle. In particular, working equations for molecules subject to intense laser fields are formulated in a very flexible manner, without restriction to the full-CI case, and nuclei can been treated as either classical or quantum particles.

In this paper, as the first step towards the numerical implementation of the general TD-MCSCF method \cite{anzaki2017fully}, we report our implementation specializing for diatomic molecules in the full-CI case. By transforming to Jacobian
coordinates and neglecting the rotational motion of the nuclei, we have only one degree of freedom for the nuclear motion. Thus nuclear orbitals can be defined as functions of internuclear distance. This treatment highly simplifies the underlying numerical procedures and at the same time retains the electron-nuclear correlation. For the electron motion, we use three-dimensional curvilinear coordinates to effectively reduce the computational cost. We apply the present implementation to a hydrogen molecule in intense laser fields. By comparing with the fixed-nuclei case, we address the importance of electron-nuclear correlation.

This paper is organized as follows. In Sec.~\ref{sec2}, we present EOMs of the TD-MCSCF method specialized to diatomic molecules. Numerical  implementation is described in Sec.~\ref{sec3}, and its applications to $\mathrm{H_2}$ and $\mathrm{D_2}$ are described in Sec.~\ref{sec4}. Section \ref{sec5} concludes this paper. Atomic units are used throughout unless otherwise stated.
\section{Theory}\label{sec2}
\subsection{The system Hamiltonian}\label{sec2a}
We consider a neutral diatomic molecule with $N$ electrons exposed to a laser field linearly polarized parallel with the molecular axis in the $z$ direction. The motions of the two nuclei [with mass and charge ($M_1$, $Z_1$) and ($M_2$, $Z_2$)] and $N$ electrons are characterized by their position vectors $\bm{R}_1$, $\bm{R}_2$, and $\{\mathbf{r}_i':i=1,2,\cdots,N$\}, respectively, in the laboratory frame. Hiskes \cite{hiskes1961dissociation} has shown that it is possible to separate the center-of-mass motion by choosing the center of mass of the two nuclei as the origin (Jacobian coordinates \cite{pack1987quantum}) and introducing $N+2$ new variables: a center-of-mass coordinate $\bm{R}_c$, a relative nuclear coordinate $\bm{R}$ and $N$ additional coordinates $\{\mathbf{r}_i:i=1,2,\cdots,N$\} representing the position vectors of the $i$-th electron from the center of mass of the two nuclei. For aligned molecules, we neglect the nuclear rotation and retain only vibrational motion. In this case, the internuclear distance vector $\bm{R}$ can be simply expressed as $\bm{R}\equiv R\bm{e}_z$, where $\bm{e}_z$ is the unit vector in the $z$ direction. After several algebraic manipulations (see e.g., Ref.~\cite{kato2009time}), the Hamiltonian for the internal motion reads
\begin{equation}\label{eq:total_Ham}
    \hat{H}(t)=\hat{H}^{(n)}(t)+\hat{H}^{(e)}(t)+\hat{H}^{(\text{ne})},
\end{equation} 
where $\hat{H}^{(n)}(t)$, $\hat{H}^{(e)}(t)$, and $\hat{H}^{\text{ne})}$ are the nuclear part, electronic part of the Hamiltonian, and electron-nuclear Coulomb interaction, respectively. $\hat{H}^{(n)}(t)$ describes the relative motion of the two nuclei with the repulsive Coulomb potential and the coupling with the laser field. Within the dipole approximation, it is given by
\begin{equation}\label{eq:LG_nHam}
    \hat{H}^{(n)}_{\text{LG}}(t)=-\frac{1}{2\mu_n}\frac{\partial^2}{\partial R^2}+\frac{Z_1Z_2}{R}-\frac{Z_1M_2-Z_2M_1}{M_1+M_2}E(t)R,
\end{equation}
in the length gauge (LG), and
\begin{equation}\label{eq:VG_nHam}
    \hat{H}^{(n)}_{\text{VG}}(t)=-\frac{1}{2\mu_n}\frac{\partial^2}{\partial R^2}+\frac{Z_1Z_2}{R}-i\frac{Z_1M_2-Z_2M_1}{M_1M_2}A(t)\frac{\partial}{\partial R},  
\end{equation}
in the velocity gauge (VG), where $\mu_n=\frac{M_1M_2}{M_1+M_2}$ is the reduced nuclear mass, and $E(t)$ and $A(t)=-\int_{-\infty}^t E(t') dt'$ are the laser electric field and vector potential, respectively. The last terms in Eq.~(\ref{eq:LG_nHam}) and (\ref{eq:VG_nHam}) vanish for homonuclear molecules such as $\mathrm{H_2}$ because there is no dipole coupling with the laser field for the nuclei. The electronic part of the Hamiltonian in Eq.~(\ref{eq:total_Ham}) reads
\begin{eqnarray}\label{eq:eHam}
    \hat{H}^{(e)}(t)&=&\sum_{i=1}^{N}\hat{h}^{(e)}(\mathbf{r}_i,t)+\sum_{i=1}^{N}\sum_{j>i}\hat{U}^{(\text{ee})}(\mathbf{r}_i,\mathbf{r}_j)\nonumber\\
    &&+\sum_{i=1}^{N}\sum_{j> i}\hat{U}^{(\text{mp})}(\mathbf{r}_i,\mathbf{r}_j),
\end{eqnarray}
with the one-body electronic Hamiltonian either in LG or in VG
\begin{equation}\label{eq:LG_seHam}
    \hat{h}^{(e)}_{\text{LG}}(\mathbf{r},t)=-\frac{1}{2\mu_e}\nabla^2+E(t)z,
\end{equation}

\begin{equation}\label{eq:VG_seHam}
    \hat{h}^{(e)}_{\text{VG}}(\mathbf{r},t)=-\frac{1}{2\mu_e}\nabla^2-i\left(1+\frac{Z_1+Z_2}{M_1+M_2}\right)A(t)\frac{\partial}{\partial z},
\end{equation}
the electron-electron Coulomb interaction
\begin{equation}\label{eq:eeHam}
    \hat{U}^{(\text{ee})}(\mathbf{r}_1,\mathbf{r}_2)=\frac{1}{\left|\mathbf{r}_1-\mathbf{r}_2\right|},
\end{equation}
and the mass-polarization term
\begin{equation}\label{eq:mp}
    \hat{U}^{(\text{mp})}(\mathbf{r}_1,\mathbf{r}_2)=\frac{1}{M_1+M_2}\bm \nabla_1\cdot\bm \nabla_2,
\end{equation}
where $\mu_e=\frac{M_1+M_2}{M_1+M_2+1}$ is the reduced electron mass. The mass-polarization term Eq.~(\ref{eq:mp}) describes the deviation of the center of mass of the nuclei from the center of mass of the nuclei plus any $N-1$ subset of the $N$ electrons, which is a minor term \cite{palacios2015theoretical} and omitted in the present paper. Finally, the last term in Eq.~(\ref{eq:total_Ham}) represents the electron-nuclear interaction, which has the form
\begin{equation}\label{eq:enHam}
    \hat{H}^{(\text{ne})}(\mathbf{r},R)=-\frac{Z_1}{\left|\mathbf{r}+\frac{M_2}{M_1+M_2}\bm{R}\right|}-\frac{Z_2}{\left|\mathbf{r}-\frac{M_1}{M_1+M_2}\bm{R}\right|}.
\end{equation}
It is worth noting that in the fixed-nulcei model, the electron-nuclear Coulomb interaction is trivially a time-independent quantity which enters to the one-body electron Hamiltonian. However, in the quantum-nuclei case, this term is a two-body operator and needs to be propagated explicitly.

\subsection{The TD-MCSCF method for the diatomic molecule}\label{sec2b}
In this subsection, we present the TD-MCSCF method \cite{anzaki2017fully} specializing for diatomics in the full-CI case. For a compact representation, the formulation of the theory is given in the second quantization formalism. Hereafter, we use orbital indices $\{\mu,\nu,\lambda,\gamma,\delta\}$ for electronic orbitals, $\{p,q,r\}$ for nuclear orbitals, and $\{i,j\}$ for the general (electronic and nuclear) occupied orbitals. For electrons, the Hamiltonian, Eq.~(\ref{eq:eHam}), can be rewritten as (the mass-polarization term is neglected)
\begin{equation}\label{eq:sq_eHam}
    \hat{H}^{(e)}=\sum_{\mu,\nu}(h_{e})^{\mu}_{\nu}(\hat{E_e})^{\mu}_{\nu}+\frac{1}{2}\sum_{\mu,\nu,\lambda,\gamma}(g_{e})^{\mu\lambda}_{\nu\gamma}(\hat{E_e})^{\mu\lambda}_{\nu\gamma},
\end{equation}
where $(\hat{E}_e)^{\mu}_{\nu}=\sum_{\sigma}\hat{a}_{\mu\sigma}^{\dag}\hat{a}_{\nu\sigma}$ and $(\hat{E}_e)^{\mu\lambda}_{\nu\gamma}=\sum_{\sigma\sigma'}\hat{a}_{\mu\sigma}^{\dag}\hat{a}_{\lambda\sigma'}^{\dag}\hat{a}_{\gamma\sigma'}\hat{a}_{\nu\sigma}$ with $\left\{\hat{a}_{\mu\sigma}^{\dag},\hat{a}_{\mu\sigma}:\sigma\in\uparrow,\downarrow\right\}$ being the fermionic electron creation and annihilation operators, which change the occupation of spin orbitals $\{\psi_{\mu}\times s(\sigma)\}$ with $M_e$  orthonormal electron spatial functions $\{\psi_{\mu}\}$. The one- and two-electron Hamiltonian matrix elements are defined as
\begin{equation}\label{eq:seMatele}
    (h_{e})^{\mu}_{\nu}=\int d\mathbf{r} \psi^{*}_{\mu}(\mathbf{r})\hat{h}^{(e)}(\mathbf{r})\psi_{\nu}(\mathbf{r}),
\end{equation}
and 
\begin{equation}\label{eq:teMatele}
    (g_{e})^{\mu\lambda}_{\nu\gamma}=\iint d\mathbf{r}_1 d\mathbf{r}_2
    \frac{\psi^{*}_{\mu}(\mathbf{r}_1)\psi^{*}_{\lambda}(\mathbf{r}_2)\psi_{\nu}(\mathbf{r}_1)\psi_{\gamma}(\mathbf{r}_2)}{\left|\mathbf{r}_1-\mathbf{r}_2\right|}.
\end{equation}
Similarly, the operators, Eq.~(\ref{eq:LG_nHam}), (\ref{eq:VG_nHam}) and (\ref{eq:enHam}), in the second quantization notation are expressed as
\begin{equation}\label{eq:sq_nHam}
    \hat{H}^{(n)}=\sum_{p,q}(h_{n})^{p}_{q}(\hat{E_n})^{p}_{q},
\end{equation}
and
\begin{equation}\label{eq:sq_enHam}
    \hat{H}^{(\text{ne})}=\sum_{p,q,\mu,\nu}(g_{\text{ne}})^{p\mu}_{q\nu}(\hat{E_n})^{p}_{q}(\hat{E_e})^{\mu}_{\nu},
\end{equation}
where $(\hat{E}_n)^{p}_{q}=\hat{b}_{p}^{\dag}\hat{b}_{q}$ and $\hat{b}_{p}^{\dag}$($\hat{b}_{p}$) is the nuclear creation (annihilation) operator for the set of $M_n$ nuclear spatial orbitals $\{\chi_{p}\}$. The matrix elements, $(h_{n})^{p}_{q}$ and $(g_{\text{ne}})^{p\mu}_{q\nu}$, are defined as
\begin{equation}\label{eq:snMatele}
    (h_n)^p_q=\int dR\chi^*_p(R)\hat{h}^{(n)}(R)\chi_q(R),
\end{equation}
and
\begin{eqnarray}\label{eq:neMatele}
    (g_{\text{ne}})^{p\mu}_{q\nu}&=&\iint dR d\mathbf{r}\chi^*_p(R)\psi^*_{\mu}(\mathbf{r})\hat{H}^{(\text{ne})}(\mathbf{r},R)\nonumber\\
    &&\times\chi_q(R)\psi_{\nu}(\mathbf{r}).
\end{eqnarray}
For the convenience of later discussion, we define the one- and two-body reduced density matrix (RDM) elements as
\begin{subequations}\label{eq:RDMs}
\begin{equation}\label{eq:1eRDM}
        (\rho_e)^{\mu}_{\nu}=\bra{\Psi}(\hat{E}_e)^{\nu}_{\mu}\ket{\Psi}, 
\end{equation}
\begin{equation}\label{eq:1nRDM}
     (\rho_n)^{p}_{q}=\bra{\Psi}(\hat{E}_n)^{q}_{p}\ket{\Psi}, 
\end{equation}
\begin{equation}\label{eq:2eRDM}
     (\rho_{\text{ee}})^{\mu\lambda}_{\nu\gamma}=\bra{\Psi}(\hat{E}_e)_{\mu\lambda}^{\nu\gamma}\ket{\Psi},
\end{equation}
\begin{equation}\label{eq:neRDM}
    (\rho_{\text{ne}})^{p\mu}_{q\nu}=\bra{\Psi}(\hat{E}_n)^{q}_{p}(\hat{E}_e)^{\nu}_{\mu}\ket{\Psi},
\end{equation}
\end{subequations}
where $\Psi$ is the total molecular wave function defined below in Eq.~(\ref{eq:ne_wfn}).

In the TD-MCSCF method limited to full-CI expansion, the total electron-nuclear wave function of a diatomic molecule is expanded as
\begin{equation}\label{eq:ne_wfn}
    \Psi(t)=\sum_{I,p}C_{I,p}(t)\Phi_I(t)\chi_p(t),
\end{equation}
where $\{\Phi_I\}$ denotes the $N$-electron Slater determinants constructed from electronic orbitals $\{\psi_{\mu}\}$, and $\{\chi_p\}$ are nuclear orbitals. 

The EOMs for the TD-MCSCF method have been derived based on the time-dependent variational principle \cite{lowdin1972some,moccia1973time}, where the action integral $S$, 
\begin{equation}\label{eq:TDVP}
    S=\int dt\bra{\Psi}\hat{H}-i\frac{\partial}{\partial t}\ket{\Psi},
\end{equation}
is made stationary with respect to any possible variations of orbitals $\{\psi_{\mu}\}$ and $\{\chi_p\}$ as well as the CI coefficients $\{C_{I,p}\}$. The detailed derivation has been given in Ref.~\cite{anzaki2017fully}. Here in our notation, the resulting EOMs of the CI coefficients read
\begin{equation}\label{eq:ci_EOM}
    i\frac{d}{dt}C_{I,p}=\sum_{J,q}\bra{\Phi_I\chi_p}\hat{H}-\hat{X}^{(n)}-\hat{X}^{(e)}\ket{\Phi_J\chi_q}C_{J,q}.
\end{equation}
The EOMs of the nuclear and electronic orbitals are given by
\begin{widetext}
\begin{equation}\label{eq:n_EOM}
i\frac{\partial}{\partial t}\ket{\chi_i}=\hat{Q}^{(n)}\left\{\hat{h}^{(n)}\ket{\chi_i}+\sum_{p,q}(\rho_n^{-1})^p_i\sum_{\mu,\nu}(\rho_{\text{ne}})_{p\mu}^{q\nu}(\hat{W}_{\text{ne}})^{\mu}_{\nu}\ket{\chi_q}\right\}+\sum_j\ket{\chi_j}(X_n)^j_i,
\end{equation}
\begin{equation}\label{eq:e_EOM}
    i\frac{\partial}{\partial t}\ket{\psi_i}=\hat{Q}^{(e)}\Bigg\{\hat{h}^{(e)}\ket{\psi_i}+\sum_{\mu,\nu}(\rho_e^{-1})^{\mu}_{i}\Big[\sum_{\lambda,\gamma}(\rho_{\text{ee}})_{\mu\lambda}^{\nu\gamma}(\hat{W}_{e})^{\lambda}_{\gamma}+\sum_{p,q}(\rho_{\text{ne}})_{p\mu}^{q\nu}(\hat{W}_{\text{en}})^{p}_{q}\Big]\ket{\psi_{\nu}}\Bigg\}+\sum_j\ket{\psi_j}(X_e)^j_i.
\end{equation}
\end{widetext}
In Eq.~(\ref{eq:ci_EOM})-(\ref{eq:e_EOM}), the operators $\hat{Q}^{(n)}=1-\sum_p\ket{\chi_p}\bra{\chi_p}$ and $\hat{Q}^{(e)}=1-\sum_{\mu}\ket{\psi_{\mu}}\bra{\psi_{\mu}}$ are projectors onto the orthogonal complement of the occupied orbital space, which guarantee the orthonormality of orbitals during time propagation. $(\hat{W}_{e})^{\mu}_{\nu}$, $(\hat{W}_{\text{ne}})^{\mu}_{\nu}$, and $(\hat{W}_{\text{en}})^{p}_{q}$ are mean-field potentials, given, in the coordinate space, by
\begin{equation}\label{eq:ee_mf}
    (\hat{W}_{e})^{\mu}_{\nu}(\mathbf{r})=\int d\mathbf{r}'\frac{\psi_{\mu}^*(\mathbf{r}')\psi_{\nu}(\mathbf{r}')}{|\mathbf{r}-\mathbf{r}'|},
\end{equation}
\begin{equation}\label{eq:ne_mf}
    (\hat{W}_{\text{ne}})^{\mu}_{\nu}(R)=\int d\mathbf{r}\psi_{\mu}^*(\mathbf{r})\hat{H}^{(\text{ne})}(\mathbf{r},R)\psi_{\nu}(\mathbf{r}),
\end{equation}
and
\begin{equation}\label{eq:en_mf}
     (\hat{W}_{\text{en}})^{p}_{q}(\mathbf{r})=\int dR\chi_{p}^*(R)\hat{H}^{(\text{ne})}(\mathbf{r},R)\chi_{q}(R).   
\end{equation}
The operators, $\hat{X}^{(n)}$ and $\hat{X}^{(e)}$, the matrix elements of which are defined by
\begin{equation}\label{eq:Xn}
    (\hat{X}_n)^p_q=i\bra{\chi_p}\frac{\partial}{\partial t}\ket{\chi_q},
\end{equation}
and
\begin{equation}\label{eq:Xe}
    (\hat{X}_e)^{\mu}_{\nu}=i\bra{\psi_{\mu}}\frac{\partial}{\partial t}\ket{\psi_{\nu}},
\end{equation}
determine the components of the time derivation of the orbitals within the occupied orbital space. These two operators impose constraints to the EOMs, which can be arbitrary Hermitian matrices. Depending on the choice of the constraint operators, the EOMs have various equivalent forms. In the present paper, we choose the natural-orbital constraint, i.e., the form of the constraint operators $\hat{X}^{(n)}$ and $\hat{X}^{(e)}$ that satisfies the condition to diagonalize the one-body RDMs for both electrons and nuclei \cite{beck2000multiconfiguration}. In other words, electronic and nuclear natural orbitals \cite{lowdin1955quantum} are propagated directly during the interaction with the laser pulse. This form is particularly suited for our propagation scheme applied below. Explicit expressions for the natural-orbital constraint operators together with their variants in imaginary-time propagation for ground-state calculation are given in Appendix \ref{app:A}. For completeness, we also briefly summarize the working equations of TD-MCSCF method for a pure multielectron system within fixed nuclei in Appendix \ref{app:B}.
\section{Implementation}\label{sec3}
One of the important aspects for the implementation of TD-MCSCF method using real-space grid-based techniques is the choice of coordinate system. Many existing implementations of TD-MCSCF method are optimized to efficiently study the strong-field phenomena in atoms and diatomic molecules in the presence of either linearly or elliptically polarized laser pulse, exploiting the underlying symmetries of the system Hamiltonian with, for instance, the spherical \cite{Sato2016,Orimo2018,orimo2019application,hochstuhl2011two}, cylindrical \cite{lotstedt2019time}, or prolate spheroidal coordinates \cite{haxton2011multiconfiguration}. To retain the possibility of carrying out first-principle simulations of the response of molecules to intense laser pulses without assuming symmetry, we do not pursue this direction here. 

The most straightforward way is using Cartesian coordinates with an equal grid mesh. While this approach offers better opportunities for parallel computing due to the highly sparse and structured representation of the Hamiltonian, it suffers from drawbacks of requiring small grid spacings to accurately represent the electron-nucleus interaction and a large number of grid points to sustain photoelectrons (or dissociated nuclei). One way to overcome this problem is using multiresolution Cartesian grids \cite{Sawada2016}, which has previously been demonstrated for the implementation of MCTDHF method. Alternatively, we can use curvilinear coordinates \cite{Gygi1995,Zumbach1996,Dundas2012}. By choosing an appropriate coordinate transformation, it is possible to achieve a high spatial resolution in the vicinity of nuclei and, at the same time, a sufficiently large simulation box. Furthermore, the coordinate transformation can be given in analytical form. In addition, the EOMs can be analytically transformed. The transformed EOMs are discretized using higher-order finite difference methods on \textit{an equal grid mesh} in  curvilinear coordinates. Such a treatment overcomes the shortcomings of the equal grid mesh while retaining its advantages. This scenario is demonstrated for the implementation of the time-dependent Hartree-Fock (TDHF) method elsewhere. In Sec.~\ref{sec3a}, we describe the coordinate transformation for the present case.

Another important issue is the time propagation method. The orbital EOMs Eq.~(\ref{eq:n_EOM}) and (\ref{eq:e_EOM}) consist of stiff linear terms and nonstiff nonlinear terms. The stiffness of the linear part comes from the one-body kinetic-energy operators while the nonlinear terms contain dependencies on the CI coefficients and the orbitals. For an efficient propagation of the EOMs, we adopt the exponential time differencing second-order Runge-Kutta scheme (ETDRK2)  \cite{kidd2017exponential,hochbruck2010exponential}, as described in Sec.~\ref{sec3b}.
\subsection{Curvilinear Coordinate}\label{sec3a}
Here we mainly focus on the electron coordinate transformation and briefly mention the nuclear transformation at the end of this subsection. Let us consider the physical space to be described by Cartesian coordinates $\mathbf{r}=(x,y,z)=(r_1,r_2,r_3)$ and an analytical transformation $\mathbf{r}=\bm{f}(\bm{\upxi})$ to curvilinear coordinates $\bm{\upxi}=(\xi_1,\xi_2,\xi_3)$ that act as the computational space. This transformation is single-valued and at least twice continuously differentiable on a three-dimensional domain. The Jacobian of the transformation $\bm{J}$ is defined as a $3\times3$ matrix of which the $(i,j)$ matrix element is  $J^i_j=\partial r_i/\partial \xi_j$, with $|J|=\text{det} \bm{J}$ being its determinants. The metric tensor $g^i_j$ in the curvilinear coordinates is 
\begin{eqnarray}\label{eq:metric_tensor}
g^i_j=\sum^3_{k=1}(J^{-1})^i_k(J^{-1})^j_k
     =\sum^3_{k=1}\frac{\partial\xi_i}{\partial r_k}\frac{\partial\xi_j}{\partial r_k},
\end{eqnarray}
and the Laplacian operator is given by
\begin{equation}\label{eq:Laplacian}
    \nabla^2=\sum^3_{i=1}\sum^3_{j=1}\frac{1}{|J|}\frac{\partial}{\partial \xi_i}|J|g^i_j\frac{\partial}{\partial \xi_j}.
\end{equation}
It is beneficial to transform the electron orbitals as
\begin{equation}\label{eq:orbtrans}
    \tilde{\psi}_{\mu}(\bm{\upxi}) = \sqrt{|J|}\psi_{\mu}(\mathbf{r}(\bm{\upxi})),
\end{equation}
such that the inner product obeys
\begin{equation}\label{eq:innerprod}
      \int d\mathbf{r} \psi^*_{\mu} (\mathbf{r})\psi_{\nu}(\mathbf{r})
  =
  \int d\bm{\upxi} \tilde{\psi}^*_{\mu} (\bm{\upxi})\tilde{\psi}_{\nu}(\bm{\upxi}).
\end{equation}
Inserting Eq.~(\ref{eq:orbtrans}) into Eq.~(\ref{eq:e_EOM}), any electron operator $\hat{h}$ in the EOMs of transformed electron orbitals will change according to $\hat{h}\longrightarrow\sqrt{|J|}\hat{h}\frac{1}{\sqrt{|J|}}$. The transformed Laplacian reads
\begin{equation}\label{eq:trans_Laplacian}
    \sqrt{|J|}\nabla^2\frac{1}{\sqrt{|J|}}=\sum^3_{i=1}\sum^3_{j=1}\frac{1}{\sqrt{|J|}}\frac{\partial}{\partial \xi_i}|J|g^i_j\frac{\partial}{\partial \xi_j}\frac{1}{\sqrt{|J|}}.
\end{equation}

Special care must be taken 
% when finite difference treatment of the
% Laplacian is used. It is important 
to keep the resulting Laplacian symmetric after applying the finite difference discretization \cite{Dundas2012}. In our implementation, we rewrite Eq.~(\ref{eq:trans_Laplacian}) as 
\begin{equation}\label{eq:trans_Laplacian2}
\sqrt{|J|}\nabla^2\frac{1}{\sqrt{|J|}} = \frac{1}{2}\sum_{i=1}^3\sum_{j=1}^3 \left[ g^i_j\frac{\partial^2}{\partial \xi_i \partial \xi_j} + \frac{\partial^2}{\partial \xi_i \partial \xi_j}g^i_j \right] + M,
\end{equation}
where
\begin{eqnarray}
  M &=& \frac{1}{4|J|^2}\left[ g^i_j\frac{\partial|J|}{\partial\xi_i}\frac{\partial|J|}{\partial\xi_j}
   -2|J|^2 \frac{\partial^2 g^i_j}{\partial \xi_i \partial \xi_j}
   \right. \nonumber \\
   &-& \left. 2|J|\frac{\partial g^i_j}{\partial \xi_i }\frac{\partial g^i_j}{\partial \xi_j }
   -2|J|g^i_j\frac{\partial^2 |J|}{\partial \xi_i \partial \xi_j}\right].
\end{eqnarray}
All the terms in Eq.~(\ref{eq:trans_Laplacian}) involving the first derivatives are now rewritten so that they only involve the second derivatives in Eq.~(\ref{eq:trans_Laplacian2}). When applying the finite difference method,  the symmetry of the resulting Laplacian is preserved. Other operators are formulated in a similar way.

The $z$ component of the transformed momentum operator is expressed as
\begin{equation}\label{eq:trans_pz}
    -i\sqrt{|J|}\frac{\partial}{\partial z}\frac{1}{\sqrt{|J|}}=-\frac{i}{2}\sum_{j=1}^3\left(\frac{\partial\xi_j}{\partial z}\frac{\partial}{\partial\xi_j}+\frac{\partial}{\partial\xi_j}\frac{\partial\xi_j}{\partial z}\right).
\end{equation}
The electron mean-field potential Eq.~(\ref{eq:ee_mf}) is evaluated by solving the corresponding Poisson equation
\begin{equation}\label{eq:poisson}
    \nabla^2(\hat{W}_e)^{\mu}_{\nu}(\mathbf{r})=-4\pi\psi_{\mu}^*(\mathbf{r})\psi_{\nu}(\mathbf{r}).
\end{equation}
Defining $(\hat{W}'_e)^{\mu}_{\nu}(\bm{\upxi})=\sqrt{|J|}(\hat{W}_e)^{\mu}_{\nu}(\mathbf{r}(\bm{\upxi}))$, in the same way as in Eq.~(\ref{eq:orbtrans}), the Poisson equation is transformed as
\begin{equation}\label{eq:transpoisson}
    \sqrt{|J|}\nabla^2\frac{1}{\sqrt{|J|}}(\hat{W}'_e)^{\mu}_{\nu}(\bm{\upxi})=-4\pi\sqrt{|J|}\tilde{\psi}_{\mu}^*(\bm{\upxi})\tilde{\psi}_{\nu}(\bm{\upxi}).
\end{equation}

With all the transformed operators available, the transformed EOMs can be discretized on the computational space $\bm{\upxi}=(\xi_1,\xi_2,\xi_3)$ by equally spaced grids. The fifth-order central finite difference is applied for spatial differential operators. To incorporate the exterior complex scaling (ECS) absorbing boundary \cite{McCurdy1991,Scrinzi2010,McCurdy2004,Telnov2013,Orimo2018}, we use an orthogonal transformation %introduced in our previous paper
\begin{equation}\label{eq:mapfun_abstract}
    r_i=f(\xi_i),\ \ i=1,2,3,
\end{equation}
which smoothly scales the real $\xi_i$ to complex $r_i$. The explicit expression for this transformation reads
\begin{eqnarray}\label{eq:mapfun_explicit}
\label{eq:transorth1}
f(\xi) =  \begin{cases}
    \Delta_t e^{i\theta}(\xi+\xi^{(1)}) - \xi^{(2)} \\ ( \xi < -\xi^{(1)}) \\
    \Delta_s \xi - \sum_{p=5}^8 b_p(\xi^{(1)}-\xi^{(0)})^{8-p}(\xi+\xi^{(0)})^p \\ ( -\xi^{(1)} \leq \xi < -\xi^{(0)}) \\
    \Delta_s \xi \\ ( -\xi^{(0)} \leq \xi \leq \xi^{(0)}) \\
    \Delta_s \xi + \sum_{p=5}^8 a_p(\xi^{(1)}-\xi^{(0)})^{8-p}(\xi-\xi^{(0)})^p \\ ( \xi^{(0)} < \xi \leq \xi^{(1)}) \\
    \Delta_t e^{i\theta}(\xi-\xi^{(1)}) + \xi^{(2)} \\ ( \xi^{(1)} < \xi)
    \end{cases}
\end{eqnarray}
where the computational space along one direction is divided into five regions: the inner region ($-\xi^{(0)} \leq \xi \leq \xi^{(0)}$), two intermediate regions ($ -\xi^{(1)} \leq \xi < -\xi^{(0)}$ and $\xi^{(0)} < \xi \leq \xi^{(1)}$), and two outer regions ($\xi < -\xi^{(1)}$ and $\xi^{(1)} < \xi$). The parameters $\Delta_s$ and $\Delta_t$ are the scaling factors in the inner and outer regions, respectively. The polynomials are used in the intermediate regions, where parameters $a_p$ and $b_p=(-1)^pa_p$  are determined to satisfy the conditions of continuity of $f(\xi)$ and its first and second derivatives at the boundaries. $\theta$ is the ECS scaling angle.

Special care should be taken for solving the Poisson equation when analytic continuation is applied for the orbitals. In this case, solving the Poisson equation in the complex-scaled region ($|\xi|>\xi^{(0)}$) turns out to be numerically unstable \cite{Orimo2018}. In the present paper, we approximately neglect the Coulomb interaction between electrons in the scaled region and adopt \textit{octupole} expansion for the calculation of the electron mean-field potential in the scaled region as well as the boundary condition of solving the Poisson equation in the unscaled region.

The nuclear coordinate is transformed in a similar way as that of electrons, i.e., $R=f(\zeta)$, together with orbital transformation $\tilde{\chi}_p(\zeta)=\sqrt{f'}\chi_p(\zeta)$. The transformed EOMs of nuclear orbitals are numerically discretized on equally spaced points in $\zeta$ and readily propagated. The same transformation function in Eq.~(\ref{eq:mapfun_explicit}) is adopted except for the condition $\zeta>0$.

\subsection{Time propagation scheme}\label{sec3b}
We rewrite the electronic orbital EOMs Eq.~(\ref{eq:e_EOM}) as a nonlinear differential equation 
\begin{equation}\label{eq:tid_diffequ}
    \frac{\partial}{\partial t}u(t)=-i\hat{h}u(t)+N[t,u(t)],
\end{equation}
where $\hat{h}$ is a stiff linear operator and $N[t,u(t)]$ a nonstiff nonlinear reminder. We first consider the stiff part to be time-independent, i.e., $\hat{h}=-\frac{1}{2\mu_e}\nabla^2$, and put the laser interaction term into the nonlinear part. Eq.~(\ref{eq:tid_diffequ}) can be integrated over a small time interval $[t_n,t_{n+1}=t_n+\Delta t]$ as
\begin{equation}\label{eq:ETD}
    u(t_{n+1})=e^{-i\hat{h}\Delta t}u(t_n)+\int_{t_n}^{t_{n+1}}dt'e^{i\hat{h}(t_{n+1}-t')}N[t',u(t')],
\end{equation}
which is exact for time-independent $\hat{h}$. This scheme is called the exponential integrator \cite{hochbruck2010exponential}. By approximating $N[t',u(t')]$ in the integrand using second-order Runge-Kutta (RK2) method, we obtain the expressions of ETDRK2 method
\begin{align}\label{eq:ETDRK2_v1_step1}
    a(t_{n})=&u(t_n)+\Delta t\varphi_1\left(-i\hat{h}(t_n)\Delta t\right)\nonumber\\
    &\times\left\{-i\hat{h}(t_n)u(t_n)+N[t_n,u(t_n)]\right\},
\end{align}
and
\begin{align}\label{eq:ETDRK2_v1_step2}
    u(t_{n+1})=&a(t_n)+\Delta t\varphi_2\left(-i\hat{h}(t_n)\Delta t\right)\nonumber\\
    &\times\Big\{N[t_{n+1},a(t_n)]-N[t_n,u(t_n)]\Big\},
\end{align}
where the $\varphi$-functions are defined as
\begin{equation}\label{eq:varphi_funs}
    \varphi_1(z)=\frac{e^z-1}{z}, \ \varphi_2(z)=\frac{e^z-1-z}{z^2}.
\end{equation}
We use the EXPOKIT package \cite{sidje1998expokit} to efficiently calculate the action of $\varphi$ functions on the operand vectors in Eq.~(\ref{eq:ETDRK2_v1_step1}) and (\ref{eq:ETDRK2_v1_step2}). The modification for time-dependent stiff operator $\hat{h}(t)$ is given in Appendix \ref{app:C}. The nuclear EOM Eq.~(\ref{eq:n_EOM}) is propagated in the same manner. For CI EOM Eq.~(\ref{eq:ci_EOM}), we treat the totality of the right-hand side as a nonlinear term and time propagation of CI reduces to the RK2 scheme.

\section{Numerical results and discussions}\label{sec4}
\subsection{Ground-state properties of \texorpdfstring{$\mathrm{H_2}$}{Lg}}\label{sec4a}
\begin{figure*}[tbhp]
 \centerline{\includegraphics[width=18cm,angle=0,clip]{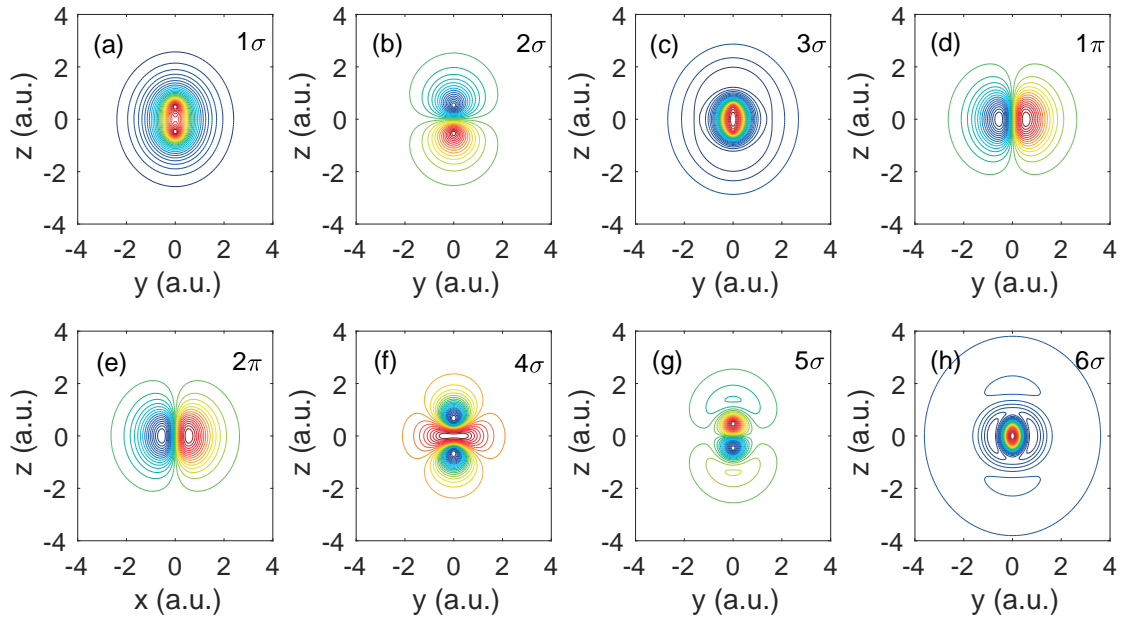}}
 \caption{Electronic natural orbitals of ground state $\mathrm{H_2}$ with fixed nuclei. The orbital space is $6\sigma2\pi$. The molecule is aligned along $z$-axis. Except for panel (e), where the orbital slice is plotted in $xz$-plane, all the other orbitals are shown in $yz$-plane.}
 \label{fig:Fig1}
\end{figure*}
To assess the performance of the method and our implementation described in Sec.~\ref{sec2} and \ref{sec3},
we first do a series of ground-state calculations taking the $\mathrm{H_2}$ molecule as an example, either with fixed-nuclei approximation or fully quantum mechanically with nuclear motion (hereafter refer to as quantum nuclei). We assume a $\mathrm{H_2}$ molecule is aligned along the $z$ axis. The imaginary-time propagation method is adopted to obtain the ground state. 
The grid resolution is optimized for ground-state calculations in order to have a better comparison with the existing literature values of ground-state properties such as ground-state energy and natural occupation number. In the quantum-nuclei case, we use $\Delta \xi=0.08$ a.u. and box size of $\xi_{\text{max}}=8$ a.u. for all the three directions of the electronic coordinates, and $\Delta \zeta=0.05$ a.u. and total length of $\zeta_{max}=6$ a.u. for nuclear coordinates. Curvilinear transformation is not used in this subsection. In the fixed-nuclei case, the same spatial grid is used for electrons, and nuclei are fixed at equilibrium internuclear distance $R=1.4$ a.u.
\begin{table}[t]
\centering
	\caption{fixed-nuclei $\mathrm{H_2}$ ground-state energies for different numbers $M$ of electron orbitals and the fraction of the covered electron-electron correlation energy. The reference energy values are taken from Ref.~\cite{Sawada2016} calculated by the MCTDHF method using a multiresolution Cartesian grid. The exact energy is taken from Ref.~\cite{guan2011breakup} by directly solving the two-electron Schrödinger equation. The equilibrium internuclear distance of $\mathrm{H_2}$ is $R=1.4$ a.u. The nuclear Coulomb repulsion energy is excluded in the ground-state energy.}
	\begin{tabularx}{\columnwidth}{XXXXX}
    \hline\hline
    \multicolumn{1}{l}{No. of}&
    \multicolumn{1}{l}{Orbital} &
    \multicolumn{2}{c}{Energy [a.u.]} &
    \multicolumn{1}{c}{Correlation} \\
    \cline{3-4}
    orbitals $M$ & space& This work &Ref.~\cite{Sawada2016}&[\%] \\
    \hline
	1 (HF) & $1\sigma$      &  -1.84792  &  -1.84123  & 0.00   \\
    2      & $2\sigma$      &  -1.86644  &  -1.85964  & 45.35  \\
    3      & $3\sigma$      &  -1.87390  &  -1.86723  & 63.61  \\
	5      & $3\sigma2\pi$  &  -1.88436  &  N/A       & 89.23  \\
	6      & $4\sigma2\pi$  &  -1.88488  &  -1.87756  & 90.50  \\
	7      & $5\sigma2\pi$  &  -1.88538  &  N/A       & 91.72  \\
	8      & $6\sigma2\pi$  &  -1.88566  &  N/A       & 92.41  \\
 	Exact  &                &  -1.88876  &            & 100.00 \\
    \hline\hline
    \end{tabularx}
    \label{tb:table1}
\end{table} 

Let us first consider the case of the fixed-nuclei $\mathrm{H_2}$ molecule. In Table \ref{tb:table1}, we present the ground-state energies for different numbers $M$ of electron orbitals. Up to eight orbitals with six $\sigma$ and two degenerate $\pi$ orbitals are used. Clearly, with increasing $M$, the ground-state energy consistently converges to the exact value \cite{guan2011breakup} obtained by the direct solution of two-electron Schrödinger equation, and only a few orbitals are needed to obtain good agreement. The energy values in this paper (third column) have a higher precision than those from Ref.~\cite{Sawada2016} (fourth column) due to finer spatial resolution. For instance, the Hartree-Fock energy presented here is nearly identical to that calculated using the finite-element discrete variable representation basis set in Ref. \cite{haxton2011multiconfiguration}.  In the fifth column, we also give the fraction of the correlation energy defined as $F_c(M)=\left[E(M)-E(1)\right]/\left[E_{\text{exact}}-E(1)\right]$. In the most accurate case $M=8$, $92.41\%$ of the correlation energy is recovered.
\begin{table}[tbp]
\centering
	\caption{Natural occupation numbers of a fixed-nuclei $\mathrm{H_2}$ molecule for $M=8$. The orbital space is $6\sigma2\pi$. The results are listed in descending order. The reference values from Ref.~\cite{haxton2011multiconfiguration} are also given for comparison. Note that there are typos for the occupation numbers of $4\sigma$ and $5\sigma$ orbitals in Ref.~\cite{haxton2011multiconfiguration}, which have been corrected in this table.}
	\begin{tabularx}{\columnwidth}{XXX}
    \hline\hline
    \multicolumn{1}{l}{}&
    \multicolumn{2}{c}{Natural occupation number} \\
    \cline{2-3}
    Orbital& This work &Ref.~\cite{haxton2011multiconfiguration} \\
    \hline
    $1\sigma$      &  1.96330                  &  1.96                   \\
    $2\sigma$      &  2.09560$\times$$10^{-2}$ &  2.10$\times$$10^{-2}$  \\
    $3\sigma$      &  6.34418$\times$$10^{-3}$ &  6.34$\times$$10^{-3}$  \\
	$1\pi$         &  4.46415$\times$$10^{-3}$ &  4.46$\times$$10^{-3}$  \\
	$2\pi$         &  4.46415$\times$$10^{-3}$ &  4.46$\times$$10^{-3}$  \\
	$4\sigma$      &  2.03040$\times$$10^{-4}$ &  2.03$\times$$10^{-4}$  \\
	$5\sigma$      &  1.90875$\times$$10^{-4}$ &  1.91$\times$$10^{-4}$  \\
	$6\sigma$      &  8.36366$\times$$10^{-5}$ &  8.36$\times$$10^{-5}$  \\
    \hline\hline
    \end{tabularx}
    \label{tb:table2}
\end{table} 
\begin{figure*}[tbhp]
 \centerline{\includegraphics[width=18cm,angle=0,clip]{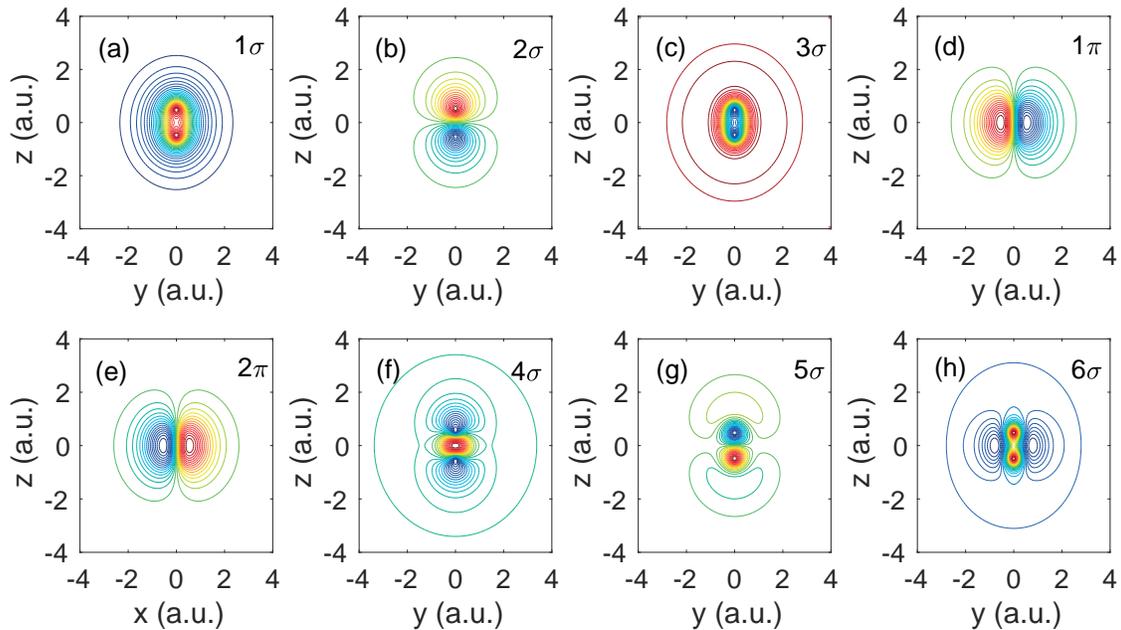}}
 \caption{Electronic natural orbitals of ground state quantum-nuclei $\mathrm{H_2}$. Same number of electronic and nuclear orbitals $M=8$ is used with electronic orbital space $6\sigma2\pi$.}
 \label{fig:Fig2}
\end{figure*}
\begin{table}[tbhp]
\centering
	\caption{Quantum-nuclei $\mathrm{H_2}$ ground-state energies for different numbers $M$ of electronic and nuclear orbitals and the fraction of the covered correlation energy. The exact energy is taken from Ref.~\cite{bubin2003variational}.}
	\begin{tabularx}{\columnwidth}{XXXXX}
    \hline\hline
    \multicolumn{1}{l}{No. of}&
    \multicolumn{1}{l}{Electronic} &
    \multicolumn{1}{l}{Energy} &
    \multicolumn{1}{l}{Correlation} \\
    orbitals $M$ &orbital space& [a.u.]&[\%] \\
    \hline
	1      & $1\sigma$      &  -1.11519  & 0.00   \\
    2      & $2\sigma$      &  -1.13434  & 39.21  \\
    3      & $3\sigma$      &  -1.14746  & 66.07  \\
	5      & $3\sigma2\pi$  &  -1.15770  & 87.04  \\
	6      & $4\sigma2\pi$  &  -1.15937  & 90.46  \\
	7      & $5\sigma2\pi$  &  -1.16014  & 92.04  \\
	8      & $6\sigma2\pi$  &  -1.16062  & 93.02  \\
 	Exact  &                &  -1.16403  & 100.00 \\
    \hline\hline
    \end{tabularx}
    \label{tb:table3}
\end{table} 

The total wave function $\Psi(t)$ of the fixed-nuclei $\mathrm{H_2}$ molecule can be expressed by using different types of electron orbitals. In the present paper, we use natural-orbital representation as explained in Appendix \ref{app:A}, where the one-body RDM is kept diagonal during the propagation. In this way, natural orbitals and natural occupation numbers, defined as eigenvectors and eigenvalues of the one-body RDM, are directly obtained after the propagation. Taking $M=8$ for example, we present the natural occupation numbers in Table \ref{tb:table2}. The natural occupation numbers represent the distribution of all electrons among natural orbitals. Their sum is equal to the number of electrons (2 for a hydrogen molecule). As can be seen, the first natural orbital is highly occupied with an occupation number close to 2 while for higher orbitals the occupation numbers decrease rapidly. The two degenerate $\pi$ orbitals have exactly the same occupation numbers, as expected. Our results are in excellent agreement with those in Ref.~\cite{haxton2011multiconfiguration} calculated by prolate spheroidal coordinate code, validating our implementation. The two-dimensional slices of natural orbitals are displayed in Fig~\ref{fig:Fig1}, where orbital symmetries can be clearly identified. The information of orbital symmetries can guide us to study the time-evolution behavior of the orbitals in the response to the laser field. For instance, the two $\pi$ orbitals in Fig~\ref{fig:Fig1}(d) and \ref{fig:Fig1}(e) will respond in the same fashion to a laser field polarized parallel to the molecular axis.

Next, we examine the ground-state properties of the quantum-nuclei $\mathrm{H_2}$ molecule. In the present paper, we always use the same number of electronic and nuclear orbitals, i.e., $M_e=M_n\equiv M$, although this is not a necessary constraint for this system \cite{beck2000multiconfiguration}. The nuclear masses of $\mathrm{H_2}$ are set to $M_1=M_2=1836.15$ a.u. The ground-state energies with respect to $M$ are shown in Table \ref{tb:table3}. Similar to the fixed-nuclei case, the ground-state energy monotonically tends to the exact value \cite{bubin2003variational} with an increasing number of orbitals. $93.02\%$ of the correlation energy is accounted at $M=8$. The nuclear and electronic natural occupation numbers for $M=8$ are presented in Table \ref{tb:table4}, with reference values from Ref.~\cite{haxton2011multiconfiguration} for comparison. The corresponding electronic natural orbitals are shown in Fig.~\ref{fig:Fig2}. We see in Fig.~\ref{fig:Fig1} and Fig.~\ref{fig:Fig2} that the overall shapes of the two sets of natural orbitals are remarkably similar, although small differences can be viewed, for instance, in the 4$\sigma$ and 6$\sigma$ orbitals. Unlike the treatment in Ref.~\cite{haxton2011multiconfiguration}, where the electron orbitals have parametric dependence on nuclear coordinate $R$ through the prolate spheroidal coordinate system, in our electron-nuclear wave function expansion Eq.~(\ref{eq:ne_wfn}), the electronic and nuclear primitive orbital functions are strictly independent without parametric dependence. The similarity of the natural orbitals in Fig.~\ref{fig:Fig1} and Fig.~\ref{fig:Fig2} thus suggests that the strong correlation between the positions of the nuclei and electrons due to their Coulomb attraction is correctly accounted for in our calculation.
\begin{table}[tbp]
\centering
	\caption{Electronic and nuclear natural occupation numbers of a quantum-nuclei $\mathrm{H_2}$ molecule for $M=8$. The electronic orbital space is $6\sigma2\pi$. The results are listed in descending order. The reference values of electronic natural orbitals are taken from Ref.~\cite{haxton2011multiconfiguration}.}
	\begin{tabularx}{\columnwidth}{XXXXX}
    \hline\hline
    \multicolumn{2}{c}{Orbital}&
    \multicolumn{3}{c}{Natural occupation number} \\
    \cline{1-2} \cline{3-5}
    Nuclear&Electronic&Nuclear&Electronic&Ref.~\cite{haxton2011multiconfiguration} \\
    \hline
    1&$1\sigma$&  9.968$\times$$10^{-1}$ &  1.953                 &  1.95                   \\
    2&$2\sigma$&  3.166$\times$$10^{-3}$ &  2.348$\times$$10^{-2}$&  2.35$\times$$10^{-2}$  \\
    3&$3\sigma$&  1.034$\times$$10^{-5}$ &  1.407$\times$$10^{-2}$&  1.46$\times$$10^{-2}$  \\
	4&$1\pi$   &  3.227$\times$$10^{-8}$ &  4.456$\times$$10^{-3}$&  4.43$\times$$10^{-3}$  \\
	5&$2\pi$   &  7.108$\times$$10^{-10}$&  4.456$\times$$10^{-3}$&  4.43$\times$$10^{-3}$  \\
	6&$4\sigma$&  7.098$\times$$10^{-11}$&  5.990$\times$$10^{-4}$&  6.18$\times$$10^{-4}$  \\
	7&$5\sigma$&  2.382$\times$$10^{-11}$&  3.496$\times$$10^{-4}$&  3.47$\times$$10^{-4}$  \\
	8&$6\sigma$&  3.575$\times$$10^{-12}$&  2.832$\times$$10^{-5}$&  1.82$\times$$10^{-4}$  \\
    \hline\hline
    \end{tabularx}
    \label{tb:table4}
\end{table} 

\subsection{Electron-nuclear dynamics of \texorpdfstring{$\mathrm{H_2}$}{Lg} in an intense laser field}\label{sec4b}
In this subsection, we present numerical applications of the implementation of our method to $\mathrm{H_2}$ subject to an intense laser pulse linearly polarized along the molecular axis in the $z$ direction, both in the fixed-nuclei and quantum-nuclei case. We adopt VG and assume the vector potential of the laser field of the following form:
\begin{equation}
    A(t)=\frac{\sqrt{I_0}}{w}\sin^2\left(\frac{\pi t}{N_0T}\right)\sin{\omega t},\ 0\leq t \leq N_0T,
\end{equation}
where $I_0$ is the peak intensity, $T$ the laser period, $\omega=2\pi/T$ the angular frequency and $N_0$ the total number of laser optical cycles.
\begin{table*}[tbp]
\centering
	\caption{Coordinate transformation parameters in Eq.~(\ref{eq:mapfun_explicit}) for electronic and nuclear coordinates}
	\begin{tabularx}{\textwidth}{XXXXXX}
    \hline\hline
    \multicolumn{1}{c}{}&
    \multicolumn{1}{l}{Nucleus}&
    \multicolumn{3}{l}{Electron} \\
    \cline{2-5}
    Parameter&$\zeta\rightarrow{R}$&$\xi_1\rightarrow{x}$&$\xi_2\rightarrow{y}$&$\xi_3\rightarrow{z}$  \\
    \hline
    $\xi^{(0)}$&20&10&10&60\\
    $\xi^{(1)}$&30&18&18&80\\
    $\xi^{(2)}$&49.15 + 6.47$i$       &33.32+5.18$i$       &33.32+5.18$i$       &118.30+12.94$i$            \\
	$a_5$&$(2.68+0.91i)\times10^{-6}$ &$(1.27+0.43i)\times10^{-5}$ &$(1.27+0.43i)\times10^{-5}$&$(2.09+0.71i)\times10^{-8}$  \\
	$a_6$&$(-5.36-1.81i)\times10^{-6}$&$(-2.56-0.86i)\times10^{-5}$&$(-2.56-0.86i)\times10^{-5}$&$(-4.19-1.42i)\times10^{-8}$ \\
	$a_7$&$(3.83+1.29i)\times10^{-6}$ &$(1.83+0.62i)\times10^{-5}$ &$(1.83+0.62i)\times10^{-5}$&$(2.99+1.01i)\times10^{-8}$  \\
	$a_8$&$(-9.57-3.24i)\times10^{-7}$&$(-4.57-1.54i)\times10^{-6}$&$(-4.57-1.54i)\times10^{-6}$&$(-7.48-2.53i)\times10^{-9}$  \\
    \hline\hline
    \end{tabularx}
    \label{tb:table5}
\end{table*} 
\begin{figure}[tbhp]
 \centerline{\includegraphics[width=9cm,angle=0,clip]{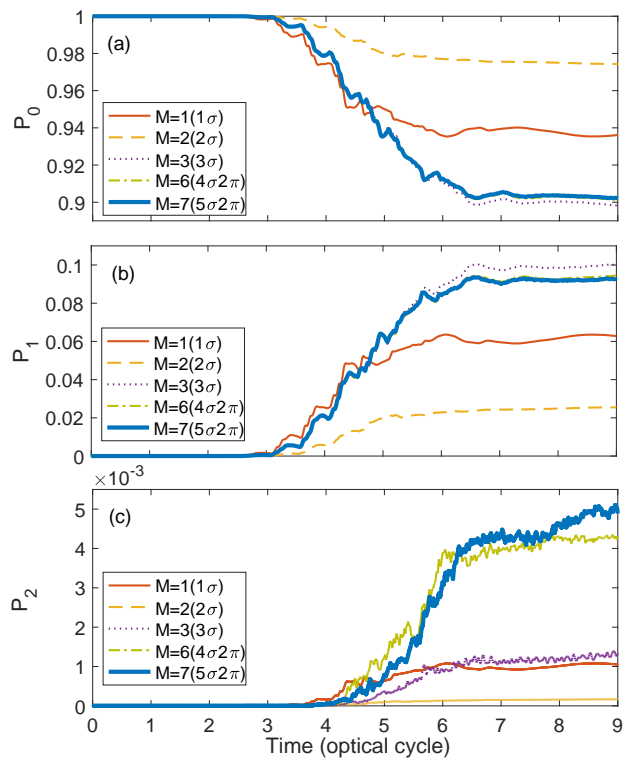}}
 \caption{Time evolution of (a) survival probability, (b) single-ionization probability, and (c) double-ionization probability of a fixed-nuclei $\mathrm{H_2}$ molecule exposed to an infrared laser pulse with a wavelength of 800 nm, an intensity of $3\times10^{14}\mathrm{W/cm^2}$ and a foot-to-foot duration of 8 optical cycles. Results of the TD-MCSCF (or MCTDHF) method are obtained with different numbers $M$ of electron orbitals.}
 \label{fig:Fig3}
\end{figure}

For time-dependent calculations, a grid width larger than that used in the ground-state calculation is employed so that the computational time falls in an affordable range. We use grid spacing $\Delta \xi=0.2$ a.u. for all the three directions of the electronic coordinates and $\Delta \zeta=0.08$ a.u. for the nuclear coordinate. The scaling factors in the curvilinear transformation Eq.~(\ref{eq:mapfun_explicit}) are chosen to be $\Delta_s=1$ and $\Delta_t=5$ and the ECS scaling angle is set to $\theta=15^{\circ}$. Other parameters in Eq.~(\ref{eq:mapfun_explicit}) for the electronic and nuclear coordinates are listed in Table \ref{tb:table5}. The ETDRK2 method described in the previous section is used to propagate EOMs with 10 000 time steps per optical cycle. We have confirmed the convergence of the results with respect to spatial and temporal resolutions.

\begin{figure}[tbhp]
 \centerline{\includegraphics[width=9cm,angle=0,clip]{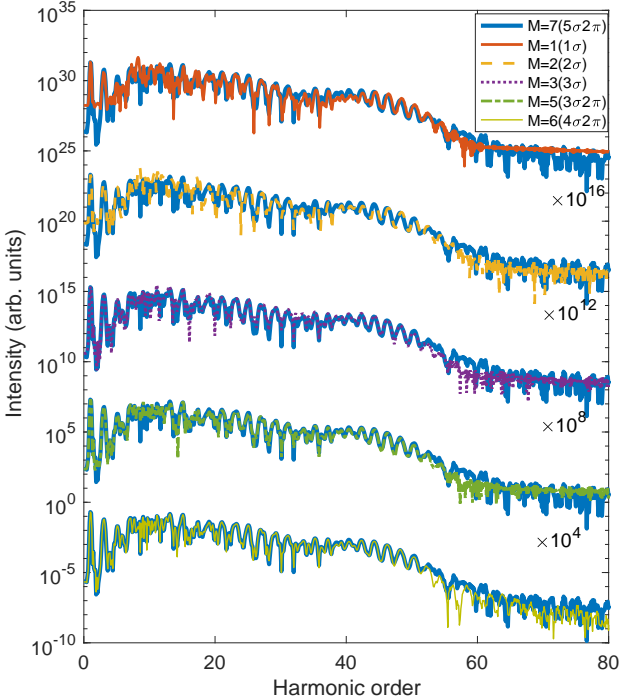}}
 \caption{High-harmonic spectra of a fixed-nuclei $\mathrm{H_2}$ molecule irradiated by an infrared laser pulse with a wavelength of 800 nm, an intensity of $3\times10^{14}\mathrm{W/cm^2}$ and a foot-to-foot duration of 8 optical cycles. Result of TD-MCSCF (or MCTDHF) method with 7 ($5\sigma+2\pi$) orbitals works as a reference. Other results with different numbers of orbitals are compared to it. The curves are vertically shifted for clarity.}
 \label{fig:Fig4}
\end{figure}

We first simulate a fixed-nuclei $\mathrm{H_2}$ molecule subject to a near infrared laser field with a wave length of 800 nm, $3\times10^{14}\mathrm{W/cm^2}$ intensity, and total duration of 8 optical cycles. For such a process, the direct exact numerical solution of the six-dimensional TDSE is still unavailable to our knowledge, due to the unfavorable scaling of the problem size with the laser wavelength. So we choose the result of the largest configuration space considered in our time-dependent simulation, $M=7$, as a reference. The simulation results are shown in Figs.~\ref{fig:Fig3} and \ref{fig:Fig4}. We present here the time evolution of the $n$-electron ionization probability (Fig.~\ref{fig:Fig3}) and the high harmonic generation (HHG) spectra (Fig.~\ref{fig:Fig4}). The $n$-electron ionization probability, $P_n$ $(n=0,1,2)$, is conveniently defined as a probability to find $n$ electrons located outside a cube with a side length of $10$ a.u, which can be evaluated using the method introduced in Ref.~\cite{Sato2013}. The HHG spectra are obtained as the modulus squared of the Fourier transform of the dipole acceleration, which, in its turn, is evaluated by double numerical differentiation of the dipole moment with respect to time. Figure~\ref{fig:Fig3} plots the survival probability $P_0$, the single-ionization probability $P_1$, and double-ionization probability $P_2$ with different number of orbitals, respectively. While $P_0$ and $P_1$ are converged for $M \geq 3$, the results with $M=1$ and $2$ show strong underestimations, which indicates the electron correlation plays a nonnegligible role during the interaction with the laser pulse. For the double ionization probability, in great contrast, we can see from Fig.~\ref{fig:Fig3}(c) that it is not fully converged even with $M=7$ and still varies after the end of the laser pulse ($t>8T_0$). This observation indicates that the double ionization is much more sensitive to the electron-electron correlation than single ionization and HHG (see below) \cite{sato2014structure,lotstedt2020excited}. The high-harmonic spectra calculated with different numbers of orbitals are shown in Fig.~\ref{fig:Fig4}. The TDHF ($M=1$) result already provides the correct cutoff position but underestimates the harmonic intensity in the plateau region. As the number of orbitals is increased, the TD-MCSCF method provides improved description of electron-electron correlation, and thus the harmonic spectrum displays a fairly rapid convergence with respect to $M$. In particular, HHG spectra with $M=6$ and $7$ show remarkable agreement with each other on the scale of the figure. 
\begin{figure}[tbhp]
 \centerline{\includegraphics[width=9.5cm,angle=0,clip]{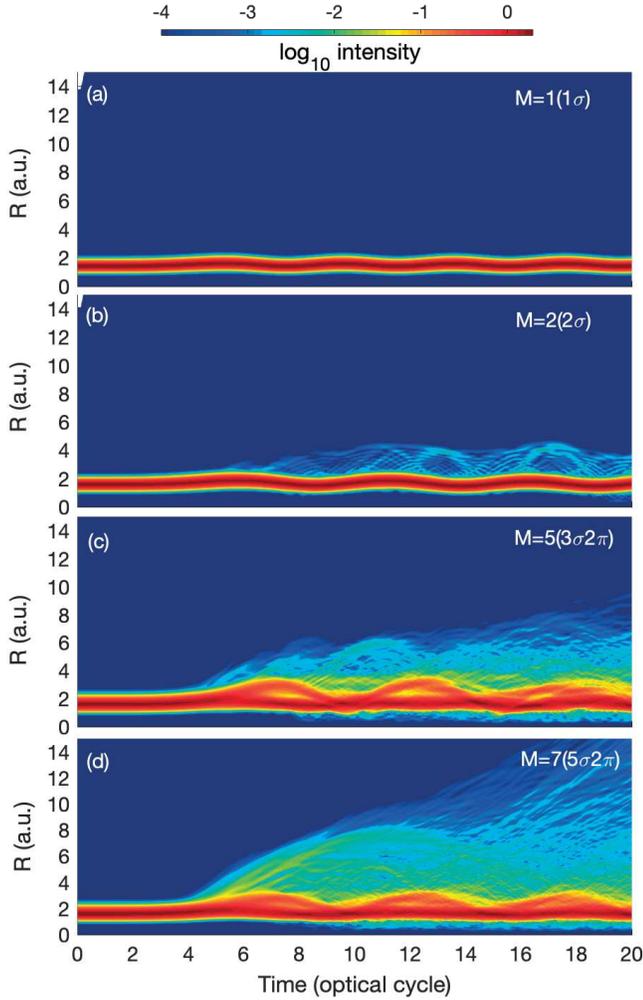}}
 \caption{Time-dependent nuclear probability density of a quantum-nuclei $\mathrm{H_2}$ molecule upon interaction with an infrared laser pulse with a wavelength of 800 nm, an intensity of $3\times10^{14}\mathrm{W/cm^2}$ and a foot-to-foot duration of 8 optical cycles. Same number of electronic and nuclear orbitals $M$ is used: (a) $M=1$, (b) $M=2$, (c) $M=5$, and (d) $M=7$.}
 \label{fig:Fig5}
\end{figure}
\begin{figure}[tbhp]
 \centerline{\includegraphics[width=9cm,angle=0,clip]{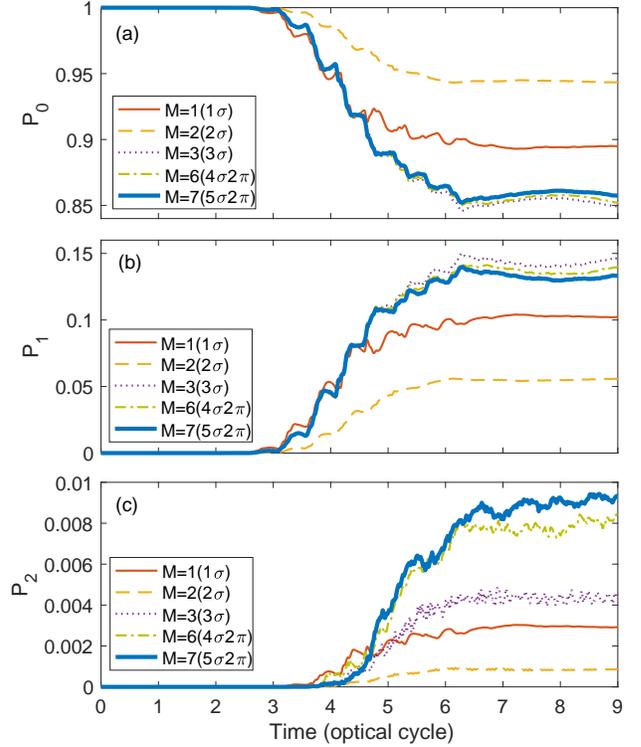}}
 \caption{Time evolution of (a) survival probability, (b) single-ionization probability, and (c) double-ionization probability of a quantum-nuclei $\mathrm{H_2}$ molecule exposed to an infrared laser pulse with a wavelength of 800 nm, an intensity of $3\times10^{14}\mathrm{W/cm^2}$ and a foot-to-foot duration of 8 optical cycles. Results of the TD-MCSCF method are obtained with different numbers of orbitals and same number of electronic and nuclear orbitals $M$ is used.}
 \label{fig:Fig6}
\end{figure}
\begin{figure}[tbhp]
 \centerline{\includegraphics[width=9cm,angle=0,clip]{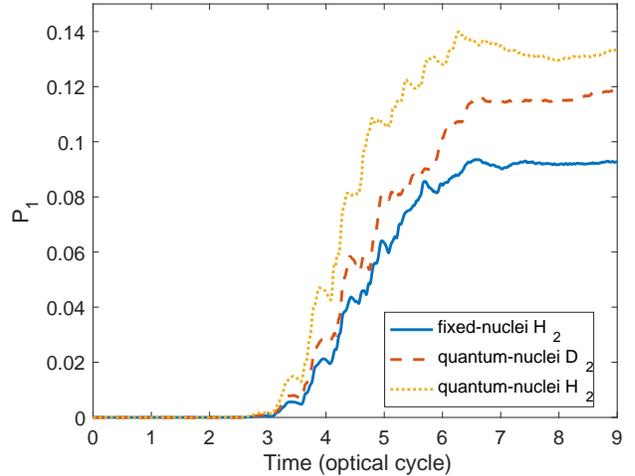}}
 \caption{Time evolution of the single-ionization probability for a fixed-nuclei $\mathrm{H_2}$ molecule, quantum-nuclei $\mathrm{D_2}$, and quantum-nuclei $\mathrm{H_2}$. Results of the TD-MCSCF method are obtained with $M=7$.}
 \label{fig:Fig7}
\end{figure}
\begin{figure}[tbhp]
 \centerline{\includegraphics[width=9cm,angle=0,clip]{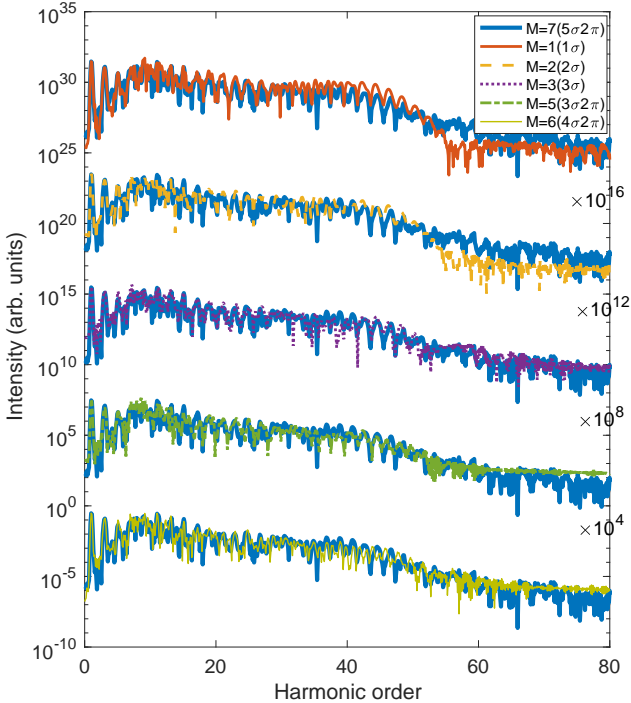}}
 \caption{High-harmonic spectra of a quantum-nuclei $\mathrm{H_2}$ molecule exposed to an infrared laser pulse with a wavelength of 800 nm, an intensity of $3\times10^{14}\mathrm{W/cm^2}$ and a foot-to-foot duration of 8 optical cycles. Same number of electronic and nuclear orbitals $M$ is used. The HHG spectra with different numbers of orbitals are compared to that with $M=7$.}
 \label{fig:Fig8}
\end{figure}

Next, let us turn to the quantum-nuclei case. The laser parameters are the same as in the fixed-nuclei case. We first analyze the time-dependent nuclear probability density, defined as the modulus squared of the total wave function integrated over the electronic coordinates. Fig~\ref{fig:Fig5} shows the logarithmically scaled, time-dependent nuclear probability density. The simulations are continued for 20 optical cycles in order to fully record the time evolution of the nuclear wave packet. In the case of $M=1$, only vibrational motion around the equilibrium internuclear distance shows up. In the TD-MCSCF calculation with $M=2$, higher order vibrational components appear. For $M=5$, even stronger vibration is seen, whereas the dissociation is still scarce. Finally, as shown in Fig.~\ref{fig:Fig5}(d), the calculation with $M=7$ leads to a much improved result: both the vibrational motions and the jetlike dissociation components are clearly observed. Our results indicate that including more orbitals leads to a better description of electron-nuclear correlation, and thus molecular vibrations as well as the dissociative ionization can be well described by the TD-MCSCF method with a sufficient number of orbitals. 

Figure~\ref{fig:Fig6} shows $n$-electron ionization probabilities of a quantum-nuclei $\mathrm{H_2}$ molecule. For this calculation, we have confirmed that the simulation box is large enough to hold the nuclear density, thus the nuclear coordinate can be safely integrated out. The results converge to that with $M=7$ (taken as a reference) as the number of orbitals increases. 
However, quantitatively, the ionization probability differs from that of the fixed-nuclei model (Fig.~\ref{fig:Fig3}) under the same laser condition. In Fig.~\ref{fig:Fig7}, we show the single ionization probability of fixed-nuclei $\mathrm{H_2}$, quantum-nuclei $\mathrm{D_2}$, and quantum-nuclei $\mathrm{H_2}$ with $M=7$. It is found that the single ionization probability of quantum-nuclei $\mathrm{H_2}$ is higher than that of $\mathrm{D_2}$, and both of them are higher than that of fixed-nuclei $\mathrm{H_2}$. This result suggests the effect of enhanced ionization due to nuclear motion \cite{dehghanian2010enhanced,chattopadhyay2019electron}.

Figure~\ref{fig:Fig8} presents the TD-MCSCF HHG spectra from the quantum-nuclei $\mathrm{H_2}$ molecule. The results show an improvement with increasing number of orbitals as in the fixed-nuclei case. However, the harmonic spectrum is not fully converged even with $M=7$. A similar trend with quantum treatment of nuclei has previously been reported for reduced-dimensional $\mathrm{H_2^+}$ \cite{anzaki2017fully}. 
Nevertheless, the quantitative agreement is already sufficient to address the effects of nuclear motion during the HHG process.
\begin{figure}[tbhp]
 \centerline{\includegraphics[width=9cm,angle=0,clip]{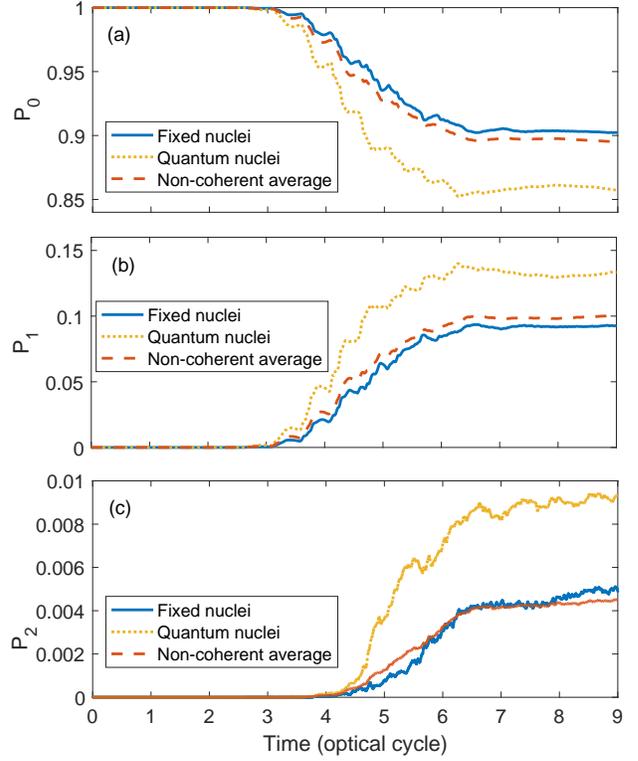}}
 \caption{(a) survival probability, (b) single-ionization probability, and (c) double-ionization probability obtained by no-coherent average of the fixed-nuclei ones are compared to the fixed-nuclei results at equilibrium internuclear distance and quantum-nuclei results. The laser parameters are the same as in Fig.~\ref{fig:Fig6}.} 
 \label{fig:Fig9}
\end{figure}
 
 While the present TD-MCSCF method treats the nuclear dynamics in a full quantum way, Saenz \textit{et al.} \cite{awasthi2005non,awasthi2010breakdown,vanne2009ionization,vanne2010alignment,forster2014ionization} have proposed to take nuclear vibration into account by the ``frozen-nuclei approximation", where the nuclear wave function is frozen to the vibrational ground state $\chi(R)$ of the electronic Born-Oppenheimer ground-state potential during the time propagation. Within this approximation, the $R$-integrated ionization probability can be calculated as the sum of fixed-nuclei signals weighted by the vibrational nuclear density \cite{forster2014ionization},
\begin{equation}
    \overline{P}(t)=\int dR P_{\text{FN}}(t;R)|\chi(R)|^2,
\end{equation}
where $P_{\text{FN}}(t;R)$ is the fixed-nuclei time-dependent ionization probability. 
We calculate $P_{\text{FN}}(t;R)$ using fixed-nuclei MCTDHF method with $M=7$ for a range of internuclear distances where the vibrational wave function $\chi(R)$ is nonvanishing, i.e., $R=1.0-2.5$ a.u. The averaged ionization probability is compared with the quantum-nuclei and fixed-nuclei results in Fig.~\ref{fig:Fig9}. On one hand, the averaged single-ionization probability is slightly higher than the fixed-nuclei one at the equilibrium internuclear distance, reflecting ionization enhanced at larger values of $R$. 
% Thus this averaging procedure can partially explain the enhanced ionization of $\mathrm{H_2}$ due to the effect of nuclear motion.  
On the other hand, quantitatively, there is a substantial discrepancy between the averaged results and those from the full quantum calculation. This observation indicates that the dynamical aspects of the electron-nuclear correlation play an important role during the interaction with the laser pulse.
% , which is absent in this frozen-nuclei approximation while fully included in our TD-MCSCF method.  

\begin{figure}[tbhp]
 \centerline{\includegraphics[width=9cm,angle=0,clip]{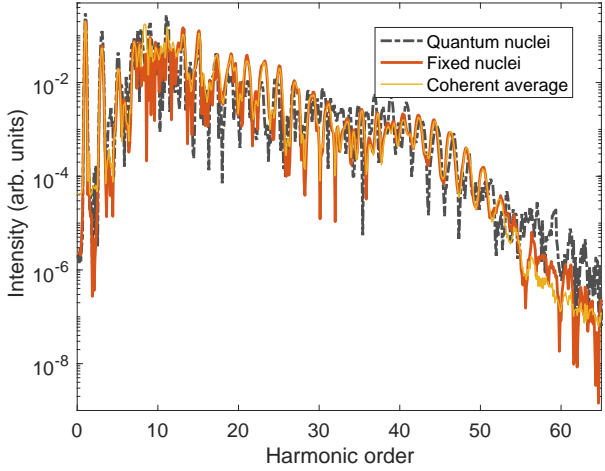}}
 \caption{The coherently averaged HHG spectrum compared to the fixed-nuclei result at equilibrium internuclear distance and quantum-nuclei result. The laser parameters are the same as in Fig.~\ref{fig:Fig8}.} 
 \label{fig:Fig10}
\end{figure}

To further examine the importance of dynamical electron-nuclear correlations, let us calculate the coherently averaged HHG spectrum.
We first obtain the time-dependent dipole acceleration averaged over internuclear distance,
\begin{equation}
    \overline{a}(t)=\int dR a_{\text{FN}}(t;R)|\chi(R)|^2,
\end{equation}
where $a_{\text{FN}}(t;R)$ denotes the fixed-nuclei dipole acceleration, calculated with the TD-MCSCF method with $M=7$. 
Then, the averaged harmonic spectrum is obtained as the modulus squared of the Fourier transform of $\overline{a}(t)$. 
The averaged spectrum is clearly different from the full quantum result and close to the fixed-nuclei result (Fig.~\ref{fig:Fig10}), due to nuclear dynamics during the quiver motion of the electron. 
% The strong electron-nuclear correlation, though largely trivial, are dynamical during the interaction with the laser pulse. 
% In our full quantum treatment, the dynamical electron-nuclear correlations are fully taken into account by explicitly propagating the corresponding two-body operator. Thus simple average over the internuclear distance can not reproduce our quantum-nuclei result. 
These results unambiguously demonstrate the importance of dynamical electron-nuclear correlations and the significance of our full quantum treatment.

\begin{figure}[tbhp]
 \centerline{\includegraphics[width=9cm,angle=0,clip]{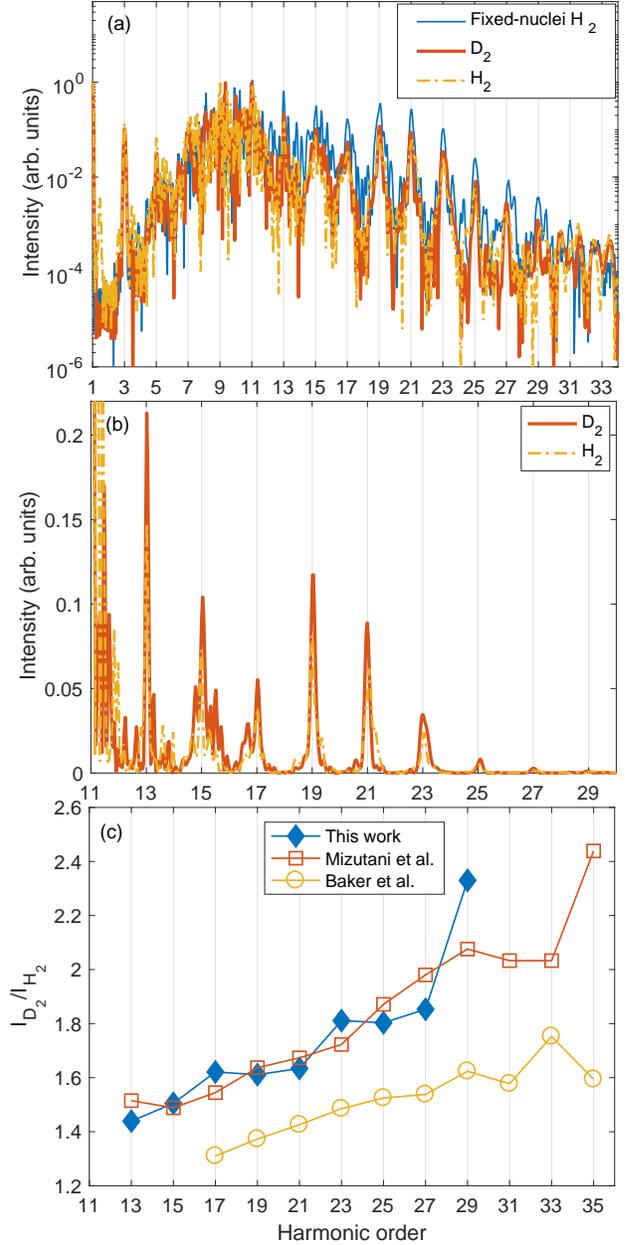}}
 \caption{(a) HHG spectra of quantum-nuclei $\mathrm{H_2}$ (yellow solid line) and $\mathrm{D_2}$ (red solid line) exposed to an infrared laser pulse with a wavelength of 800 nm, an intensity of $2\times10^{14}\mathrm{W/cm^2}$ and a foot-to-foot duration of 30 optical cycles. The HHG spectrum of fixed-nuclei $\mathrm{H_2}$ model (blue light line) is also presented for comparison. All of the results are calculated by TD-MCSCF method with $M=7$. In the calculation of $\mathrm{D_2}$, the nuclear mass is set to 3670.48 a.u. (b) HHG spectra in the plateau region of quantum-nuclei $\mathrm{H_2}$ and $\mathrm{D_2}$, plotted on a linear scale. (c) The intensity ratios of $I_{D_2}/I_{H_2}$ as a function of harmonic order. Experimental data taken from Ref.~\cite{baker2006probing} (yellow open circles) and Ref.~\cite{mizutani2011effect} (red open squares) are also included.} 
 \label{fig:Fig11}
\end{figure}

Experimentally, it has been observed that high-harmonic emission is stronger in $\mathrm{D_2}$ than that in $\mathrm{H_2}$ due to the faster nuclear motion in the lighter $\mathrm{H_2}$ molecule \cite{baker2006probing,mizutani2011effect}. We will show that the present TD-MCSCF method, which allows full quantum treatment of nuclei, can well reproduce the experimental results. We consider an 800-nm laser field with a peak intensity of $2\times10^{14}\mathrm{W/cm^2}$. The laser parameters are similar to those in the experiments. In order to reveal the odd order harmonic peaks, we use a multi-cycle laser pulse with total duration of 30 periods. The HHG spectra of quantum-nuclei $\mathrm{H_2}$ and $\mathrm{D_2}$ are calculated with $M=7$, as shown in Fig.~\ref{fig:Fig11}(a). Up to the 29th harmonics can be well resolved in the spectra. In Fig.~\ref{fig:Fig11}(b), the spectra of $\mathrm{H_2}$ and $\mathrm{D_2}$ in the plateau region are plotted on a linear scale. It can be clearly seen that the harmonic signal is stronger in $\mathrm{D_2}$ than that in $\mathrm{H_2}$. Fig.~\ref{fig:Fig11}(c) shows the intensity ratio between $\mathrm{D_2}$ and $\mathrm{H_2}$ as a function of harmonic order. The ratio is calculated by using the peak values of odd order harmonics. The ratio is larger than unity at all orders and increases with the harmonic order. Our calculation shows quantitative agreement with the experimental measurement in Ref.~\cite{mizutani2011effect} under the same laser condition. The experimental result from Ref.~\cite{baker2006probing} using a few-cycle laser pulse also exhibits a similar trend, albeit with relatively small values compared to our calculation possibly because of the difference in the laser parameters. The present calculation using the TD-MCSCF method nicely reproduces the experimentally observed isotope effects on HHG, emphasizing the importance of dynamical electron-nuclear correlation during HHG process.
\section{Summary}\label{sec5}
We have presented the numerical implementation of the recently formulated TD-MCSCF method \cite{anzaki2017fully} for diatomic molecules in the full-CI expansion. By full quantum treatment of the nuclei, the present implementation allows one to investigate coupled electron–nuclear dynamics of molecules subject to intense laser fields. As a first numerical test, we have applied this method to the ground state as well as the laser-driven dynamics of $\mathrm{H_2}$ molecule. The ground-state properties, including ground-state energy, natural orbitals and natural occupation numbers show remarkable agreement with benchmark data \cite{haxton2011multiconfiguration} available in the literature, which indicates the correctness of our implementation. For laser-driven dynamics, we have shown that the method can systematically improve the accuracy for describing the highly nonlinear HHG phenomenon by increasing the number of electronic and nuclear orbital functions. Furthermore, the TD-MCSCF method with sufficiently large number of orbitals has been applied to study the isotopic effects of HHG to justify experimental observations. Our results indicate that HHG is sensitive to the laser-induced nuclear vibrational motion. The harmonic emission is more intense in heavier isotopes and the role of nuclear motion in suppressing the intensity of high harmonics is increased with increasing harmonic order. 

While the present paper has been mainly focused on the electron–nuclear dynamics in HHG, we have also shown that the TD-MCSCF method is capable of describing photoionization and dissociation \cite{lotstedt2019time,martin2007single,cattaneo2018attosecond} of molecules interacting with a laser field. In order to properly explain experimental results such as electron-ion coincidence measurements \cite{dorner2000cold,ullrich2003recoil} during molecular dissociation, extending the functionality of the present implementation to calculate momentum distribution of ionized electrons and the joint electron–nuclear-energy spectrum will be the further aspects.
%%%%%%%%%%%%%%%%%%%%%%%%%%%%%%%%%%%%%%%%%%%%%%%%%%%%%%
\section*{acknowledgments}
{Y.~L. acknowledges Erik Lötstedt for valuable discussions on propagation methods. This research was supported in part by a Grant-in-Aid for
Scientific Research (Grants No. JP18H03891 and No. JP19H00869) from the Ministry of Education, Culture,
Sports, Science and Technology (MEXT) of Japan.
This research was also partially supported by JST COI (Grant No.~JPMJCE1313), JST CREST (Grant No.~JPMJCR15N1),
and by MEXT Quantum Leap Flagship Program (MEXT Q-LEAP) Grant Number JPMXS0118067246.}
Y.~L. gratefully acknowledges support from JSPS Postdoctoral Fellowships for Research in Japan.
%%%%%%%%%%%%%%%%%%%%%%%%%%%%%%%%%%%%%%%%%%%%%%%%%%%%%%
\appendix
\section{Derivation of natural-orbital constraints} \label{app:A}
In the TD-MCSCF method, a typical feature is the invariance of the total wave function with respect to time-dependent unitary transformations among the subspace spanned by the occupied orbitals. These unitary transformations give rise to the constraint operators, i.e., $\hat{X}^{(n)}$ and $\hat{X}^{(e)}$, appearing in the EOMs. The natural-orbital representation is one special case, where the one-body RDMs are kept to be diagonal during the time propagation. Taking $\hat{X}^{(e)}$ for example, the natural-orbital constraint matrix can be derived by considering the time derivative of the one-body RDM for electrons:
\begin{align}\label{eq:diff_rhoe_1}
    \frac{d}{dt}(\rho_e)^{\mu}_{\nu}&=&\sum_{I,p}\frac{dC^*_{I,p}}{dt}\bra{\Phi_I\chi_p}(\hat{E_e})^{\nu}_{\mu}\ket{\Psi} \nonumber \\
    &+&\sum_{I,p}\bra{\Psi}(\hat{E_e})^{\nu}_{\mu}\ket{\Phi_I\chi_p}\frac{dC_{I,p}}{dt}.
\end{align}
Substituting the CI EOMs Eq.~(\ref{eq:ci_EOM}) into Eq.~(\ref{eq:diff_rhoe_1}) and using the Hermiticity of $\hat{X}^{(e)}$ and $\hat{X}^{(n)}$, we get
\begin{align}\label{eq:diff_rhoe_2}
    \frac{d}{dt}(\rho_e)^{\mu}_{\nu}&=\sum_{I,p}i\bra{\Psi}\hat{H}-\hat{X}\ket{\Phi_I\chi_p}\bra{\Phi_I\chi_p}(\hat{E_e})^{\nu}_{\mu}\ket{\Psi} \nonumber \\
    &-\sum_{I,p}i\bra{\Psi}(\hat{E_e})^{\nu}_{\mu}\ket{\Phi_I\chi_p}\bra{\Phi_I\chi_p}\hat{H}-\hat{X}\ket{\Psi},
\end{align}
where notation $\hat{X}=\hat{X}^{(n)}+\hat{X}^{(e)}$ is used. In the case of full-CI expansion, the configuration projector $\sum_{I,p}\ket{\Phi_I\chi_p}\bra{\Phi_I\chi_p}$ can be omitted. The time derivative of the one-body RDM can be rewritten in a compact form employing the commutator $[\hat{a},\hat{b}]=\hat{a}\hat{b}-\hat{b}\hat{a}$,
\begin{eqnarray}\label{eq:diff_rhoe_3}
    \frac{d}{dt}(\rho_e)^{\mu}_{\nu}&=&-i\bra{\Psi}[(\hat{E_e})^{\nu}_{\mu},\hat{H}-\hat{X}]\ket{\Psi}\nonumber \\
    &=&-i\bra{\Psi}[(\hat{E_e})^{\nu}_{\mu},\hat{H}-\hat{X}^{(e)}]\ket{\Psi}.
\end{eqnarray}
In the natural-orbital representation, $d(\rho_e)^{\mu}_{\nu}/dt=0$ for $\mu\neq\nu$, and consequently, 
\begin{equation}\label{eq:Xe_relation}
    \bra{\Psi}[(\hat{E_e})^{\nu}_{\mu},\hat{X}^{(e)}]\ket{\Psi}=\bra{\Psi}[(\hat{E_e})^{\nu}_{\mu},\hat{H}]\ket{\Psi}.
\end{equation}
After some algebraic manipulations to evaluate the commutators, we have
\begin{equation}\label{eq:Xe_Matele}
    (X_e)^{\mu}_{\nu}=\left[(F_e)^{\mu}_{\nu}-(F_e^*)^{\nu}_{\mu}\right]\frac{(\rho_e)^{\mu}_{\mu}-(\rho_e)^{\nu}_{\nu}}{\left[(\rho_e)^{\mu}_{\mu}-(\rho_e)^{\nu}_{\nu}\right]^2+\epsilon^2},
\end{equation}
with
\begin{eqnarray}\label{eq:focke}
    (F_e)^{\mu}_{\nu}&=&\sum_{\lambda}(h_e)^{\mu}_{\lambda}(\rho_e)^{\lambda}_{\nu}+\sum_{\lambda p q}(g_{\text{ne}})^{p\mu}_{q\lambda}(\rho_{\text{ne}})_{p\nu}^{q\lambda} \nonumber \\
    &&+\sum_{\lambda \gamma \delta}(g_{\text{ee}})^{\mu\gamma}_{\lambda\delta}(\rho_{\text{ee}})_{\nu\gamma}^{\lambda\delta},
\end{eqnarray}
for off-diagonal matrix elements $\mu\neq\nu$, where $\epsilon$ is a small regularization parameter. The constraint does not fix the values of diagonal elements of operator $\hat{X}^{(e)}$, which can be simply set to zero. The same procedures can be applied to the nuclear constraint operators $\hat{X}^{(n)}$, the results are
\begin{equation}\label{eq:Xn_Matele}
        (X_n)^{p}_{q}=\left[(F_n)^{p}_{q}-(F_n^*)^{q}_{p}\right]\frac{(\rho_n)^{p}_{p}-(\rho_n)^{q}_{q}}{\left[(\rho_n)^{p}_{p}-(\rho_n)^{q}_{q}\right]^2+\epsilon^2},
\end{equation}
with
\begin{equation}\label{eq:fockn}
    (F_n)^{p}_{q}=\sum_{r}(h_n)^{p}_{r}(\rho_n)^{r}_{q}+\sum_{r \mu \nu}(g_{\text{ne}})^{p\mu}_{r\nu}(\rho_{\text{ne}})_{r\nu}^{q\mu},
\end{equation}
for $p\neq q$.

One should note that the above derivations are only valid for real-time propagation. There are differences for the presence of constraint operators between real- and imaginary-time propagation. While the constraint operators are Hermitian in real-time EOMs, they are anti-Hermitian in the imaginary time ones. In the following, we derive the expressions of constraint operators which are suitable for imaginary-time propagation. We use notations $\hat{Y}^{(e)}$ and $\hat{Y}^{(n)}$ as the imaginary-time counterparts for $\hat{X}^{(e)}$ and $\hat{X}^{(n)}$, respectively. By employing a pure imaginary time $t=-i\tau$ with $\tau$ being a real variable, the matrix elements of operators $\hat{Y}^{(e)}$ and $\hat{Y}^{(n)}$ are defined as
\begin{equation}\label{eq:Yn}
    (\hat{Y}_n)^p_q=-\bra{\chi_p}\frac{\partial}{\partial \tau}\ket{\chi_q},
\end{equation}
and
\begin{equation}\label{eq:Ye}
    (\hat{Y}_e)^{\mu}_{\nu}=-\bra{\psi_{\mu}}\frac{\partial }{\partial \tau}\ket{\psi_{\nu}}.
\end{equation}
Based on the time-dependent variation principle in imaginary time, we obtain the same EOMs for both CI coefficients and orbitals [Eq.~(\ref{eq:ci_EOM})-(\ref{eq:e_EOM})], except for a replacement $i\frac{\partial}{\partial t}\rightarrow-\frac{\partial}{\partial \tau}$ on the left-hand side of these equations. For instance, the imaginary-time CI EOMs Eq. (\ref{eq:ci_EOM}) read
\begin{equation}\label{eq:ci_EOM_imag}
    -\frac{d}{d\tau}C_{I,p}=\sum_{J,q}\bra{\Phi_I\chi_p}\hat{H}-\hat{Y}^{(n)}-\hat{Y}^{(e)}\ket{\Phi_J\chi_q}C_{J,q}.
\end{equation}
Utilizing the anti-Hermiticity of the constraint operators, the complex conjugate of Eq.~(\ref{eq:ci_EOM_imag}) is obtained as
\begin{equation}\label{eq:ci_EOM_imag_congj}
    -\frac{d}{d\tau}C^*_{I,p}=\sum_{J,q}\bra{\Phi_J\chi_q}\hat{H}+\hat{Y}^{(n)}+\hat{Y}^{(e)}\ket{\Phi_I\chi_p}C^*_{J,q}.
\end{equation}
Substituting Eq.~(\ref{eq:ci_EOM_imag}) and (\ref{eq:ci_EOM_imag_congj}) into the time derivative of electronic one-body RDM Eq.~(\ref{eq:diff_rhoe_1}) and setting the off-diagonal matrix elements to zero, we obtain a relation for $\hat{Y}^{(e)}$ as
\begin{equation}\label{eq:Ye_relation}
    \bra{\Psi}[(\hat{E_e})^{\nu}_{\mu},\hat{Y}^{(e)}]\ket{\Psi}=\bra{\Psi}\{(\hat{E_e})^{\nu}_{\mu},\hat{H}\}\ket{\Psi},
\end{equation}
where the anti-commutator $\{\hat{a},\hat{b}\}=\hat{a}\hat{b}+\hat{b}\hat{a}$ is used. Similarly, for nuclear constraint operator $\hat{Y}^{(n)}$, we have
\begin{equation}\label{eq:Yn_relation}
    \bra{\Psi}[(\hat{E_n})^{q}_{p},\hat{Y}^{(n)}]\ket{\Psi}=\bra{\Psi}\{(\hat{E_n})^{q}_{p},\hat{H}\}\ket{\Psi}.
\end{equation}
It is thus interesting to compare Eq.~(\ref{eq:Ye_relation}) for the imaginary-time propagation with Eq.~(\ref{eq:Xe_relation}) for real-time propagation: the only difference is the replacement of the commutator to the anti-commutator on the right-hand side of these two equations. The explicit expressions of $\hat{Y}^{(e)}$ $\hat{Y}^{(n)}$ involve the third-order RDMs, which are tedious and omitted here. The computation of third-order RDMs constitutes a bottleneck for numerical calculations, which needs further investigations. Nevertheless, for a hydrogen molecule with only two electrons and one nuclear degree of freedom used in the present paper, the computational effort is affordable.
\section{The TD-MCSCF formalism for fixed nuclei} \label{app:B}
In this appendix, we briefly describe the TD-MCSCF method for a multielectron system within the fixed-nuclei model. In the full-CI case, it is equivalent to MCTDHF method. We consider an $N$-electron diatomic molecule with fixed nuclei in an external laser field. The dynamics of the electronic system is described by the time-dependent Hamiltonian 
\begin{equation}
    \hat{H}(t)=\hat{H}_1(t)+\hat{H}_2,
\end{equation}
with $\hat{H}_1=\sum_i^N\hat{h}(\mathbf{r}_i,t)$ and
\begin{equation}
    \hat{h}(\mathbf{r},t)=\left(-\frac{\nabla^2}{2}-\sum_{a=1}^2\frac{Z_a}{|\mathbf{r}-\mathbf{R}_a|}+\hat{V}_{\text{ext}}\right),
\end{equation}
\begin{equation}
    \hat{H}_2=\sum_{i=1}^{N}\sum_{j>i}\frac{1}{|\mathbf{r}_i-\mathbf{r}_j|},
\end{equation}
where laser-electron interaction $\hat{V}_{\text{ext}}$  within the dipole approximation either in the LG or in the VG is given by
\begin{equation}
    \hat{V}_{\text{ext}}^{\text{LG}}(\mathbf{r},t)=\mathbf{E}(t)\cdot\mathbf{r},
\end{equation}
\begin{equation}
    \hat{V}_{\text{ext}}^{\text{VG}}(\mathbf{r},t)=-i\mathbf{A}(t)\cdot\bm \nabla.
\end{equation}
In the TD-MCSCF method, the $N$-electron wave function is expressed by a linear combination of Slater determinants as
\begin{equation}\label{eq:totewfn}
    \Psi(t)=\sum_I C_I(t)\Phi(t),
\end{equation}
where the Slater determinant $\Phi_I(t)$ is built from $M$ occupied orbitals $\{\psi_{\mu}(t)\}$. Both the CI coefficients $\{C_I\}$ and the occupied orbitals are time-dependent. By use of time-dependent  variational principle, the EOMs are derived to describe the time evolution. The resulting EOMs for the CI coefficients read
\begin{equation}
    i\frac{d}{dt}C_I(t)=\sum_J\bra{\Psi_I}\hat{H}-\hat{X}\ket{\Psi_J}C_J(t).
\end{equation}
The EOMs for the orbitals are given by
\begin{eqnarray}
    i\frac{\partial}{\partial t}\ket{\psi_i}&=&\hat{Q}\left\{\hat{h}\ket{\psi_i}+\sum_{\mu,\nu,\lambda,\gamma}(D^{-1})^{\mu}_{i}P_{\mu\lambda}^{\nu\gamma}\hat{W}^{\lambda}_{\gamma}\ket{\psi_{\nu}}\right\}\nonumber\\
    &&+\sum_j\ket{\psi_j}(X)^j_i,
\end{eqnarray}
where $\hat{Q}=1-\sum_{\mu}\ket{\psi_{\mu}}\bra{\psi_{\mu}}$ is the projector against the occupied orbital space, $D$ and $P$ the one- and two-electron RDMs, $\hat{W}^{\mu}_{\nu}$ the mean-field potential with the same definition as in Eq.~(\ref{eq:ee_mf}), and $\hat{X}$ the constraint operator defined in Eq.~(\ref{eq:Xe}). Here we also use the natural-orbital constraint $\hat{X}$ as derived in Appendix \ref{app:A}.
\section{Modification of ETDRK2 method with a time-dependent stiff part}
\label{app:C}
Following Ref.~\cite{hochbruck2010exponential}, it is possible to generalize the ETD method to treat
\begin{equation}\label{eq:td_diffequ}
    \frac{\partial}{\partial t}u(t)=-i\hat{h}(t)u(t)+N[t,u(t)],
\end{equation}
where the operator $\hat{h}(t)$ is time dependent and chosen to be the one-electron Hamiltonian $\hat{h}(t)=\hat{h}^{(e)}(t)$. Within the time interval $[t_n,t_{n+1}]$, we rewrite the operator  $\hat{h}(t)$ as
\begin{align}\label{eq:aul_seHam}
    \hat{h}(t)=&\hat{h}(t_n)+\Dot{\hat{h}}(t_n)(t-t_n)\nonumber \\
    &+\left[\hat{h}(t)-\hat{h}(t_n)\right]-\Dot{\hat{h}}(t_n)(t-t_n),
\end{align}
where $\Dot{\hat{h}}(t)=\partial \hat{h}(t)/\partial t$. By introducing an auxiliary vector $\mathcal{U}(t)=\left[t-t_n,u(t)\right]^T$ and employing Eq.~(\ref{eq:aul_seHam}), we can rewrite Eq.~(\ref{eq:td_diffequ}) coupled with $\Dot{t}=1$ as
\begin{equation}\label{eq:td_diffequ_trans}
        \frac{\partial}{\partial t}\mathcal{U}(t)=-i\mathcal{H}\mathcal{U}(t)+\mathcal{N}(t),
\end{equation}
with
\begin{equation}\label{eq:HandN}
    \mathcal{H}=
    \begin{bmatrix}
    0&0\\ \Dot{\hat{h}}(t_n)u(t_n)&\hat{h}(t_n)
    \end{bmatrix}, \ 
        \mathcal{N}(t)=
    \begin{bmatrix}
    1 \\ N'[t,u(t)]
    \end{bmatrix},
\end{equation}
where
\begin{align}\label{eq:Nprime}
    N'[t,u(t)]=&N[t,u(t)]-i\left[\hat{h}(t)-\hat{h}(t_n)\right]u(t)\nonumber\\
    &+i\Dot{\hat{h}}(t_n)(t-t_n)u(t_n).
\end{align}
It is obvious that $\mathcal{H}$ is now time independent, thus the ETD method with the time-independent linear operator can be applied to Eq.~(\ref{eq:td_diffequ_trans}). The $\varphi$ functions of $\mathcal{H}$ are give by
\begin{widetext}
\begin{equation}\label{eq:Htrans}
    \varphi_k(-i\mathcal{H}\Delta t)=
    \begin{bmatrix}
    \varphi_{k}(0)&0 \\
    -i\Delta t\varphi_{k+1}(-i\hat{h}(t_n)\Delta t)\Dot{\hat{h}}(t_n)u(t_n)&\varphi_k(-i\hat{h}(t_n)\Delta t)
    \end{bmatrix}.
\end{equation}
The final expressions for ETDRK2 method are
\begin{align}
    a(t_n)=u(t_n)+\Delta t \varphi_1(-i\hat{h}(t_n)\Delta t) \left \{N[t_n,u(t_n)]-i\hat{h}(t_n)u(t_n)\right \}-i\Delta t^2 \varphi_2(-i\hat{h}(t_n)\Delta t)\Dot{\hat{h}}(t_n)u(t_n),
\end{align}
and
\begin{align}
    u(t_{n+1})=a(t_n)+\Delta t\varphi_2(-i\hat{h}(t_n)\Delta t)\left\{W[t_{n+1},a(t_n)]-W[t_n,u(t_n)]-i\{\hat{h}(t_{n+1})-\hat{h}(t_n)\}a(t_n)+i\Delta t\Dot{\hat{h}}(t_n)u(t_n)\right\}.
\end{align}
\end{widetext}
%%%%%%%%%%%%%%%%%%%%%%%%%%%%%%%%%%%%%%%%%%%%%%%%%%%%%%
\bibliography{ref}

%%%%%%%%%%%%%%%%%%%%%%%%%%%%%%%%%%%%%%%%%%%%%%%%%%%%%%
\end{document}